\documentclass[prb,a4paper,eqsecnum,aps,showpacs,floatfix,twocolumn,byrevtex]{revtex4}
\usepackage{graphicx}
\begin{document}
\title{Magnetic and thermal properties of 4f--3d ladder-type molecular compounds}
\author{M. Evangelisti\cite{byline}}
\affiliation{Kamerlingh Onnes Laboratory, Leiden University, 2300 RA, Leiden, The Netherlands and\\
Instituto de Ciencia de Materiales de Arag\'on, CSIC-Universidad de Zaragoza, 50009 Zaragoza,
Spain}
\author{M. L. Kahn\cite{byline2}}
\affiliation{Laboratoire des Sciences Mol\'eculaires, Institut de Chimie de la Mati\`ere
Condens\'ee de Bordeaux, UPR CNRS 9048, 33608 Pessac, France}
\author{J. Bartolom\'e}
\affiliation{Instituto de Ciencia de Materiales de Arag\'on, CSIC-Universidad de Zaragoza, 50009
Zaragoza, Spain}
\author{L. J. de Jongh}
\affiliation{Kamerlingh Onnes Laboratory, Leiden University, 2300 RA, Leiden, The Netherlands}
\author{C. Meyers}
\author{J. Leandri}
\author{Y. Leroyer}
\affiliation{Centre de Physique Mol\'eculaire Optique et Hertzienne, Universit\'e Bordeaux 1, UMR
CNRS 5798, 33405 Talence, France}
\author{C. Mathoni\`ere}
\affiliation{Laboratoire des Sciences Mol\'eculaires, Institut de Chimie de la Mati\`ere
Condens\'ee de Bordeaux, UPR CNRS 9048, 33608 Pessac, France }
\date{\today}
\begin{abstract}
We report on the low-temperature magnetic susceptibilities and specific heats of the isostructural
spin-ladder molecular complexes L$_{2}$[M(opba)]$_{3}\cdot x$DMSO$\cdot y$H$_{2}$O, hereafter
abbreviated with L$_{2}$M$_{3}$ (where L~=~La, Gd, Tb, Dy, Ho and M~=~Cu, Zn). The results show
that the Cu containing complexes (with the exception of La$_{2}$Cu$_{3}$) undergo long-range
magnetic order at temperatures below 2~K, and that for Gd$_{2}$Cu$_{3}$ this ordering is
ferromagnetic, whereas for Tb$_{2}$Cu$_{3}$ and Dy$_{2}$Cu$_{3}$ it is probably antiferromagnetic.
The susceptibilities and specific heats of Tb$_{2}$Cu$_{3}$ and Dy$_{2}$Cu$_{3}$ above $T_{C}$ have
been explained by means of a model taking into account nearest as well as next-nearest neighbor
magnetic interactions. We show that the intraladder L--Cu interaction is the predominant one and
that it is ferromagnetic for L~=~Gd, Tb and Dy. For the cases of Tb, Dy and Ho containing
complexes, strong crystal field effects on the magnetic and thermal properties have to be taken
into account. The magnetic coupling between the (ferromagnetic) ladders is found to be very weak
and is probably of dipolar origin.
\end{abstract}


\maketitle

\section{Introduction}

In today's search for smaller, faster, more selective and efficient products and processes, the
engineering of well-defined spatial microarrangements of pure and composite materials is of vital
importance for the creation of new magnetic devices. A possibility to assemble microstructures in a
controlled way is to use molecular-based materials. The design of novel ferromagnetic molecular
materials is certainly among the stimulating subjects.~\cite{6gen} In such a context, molecular
complexes based on transition metal ions are good candidates because the procedures for predicting
the ferromagnetic nature of the interaction in this class of materials have become fairly well
established.~\cite{6kahn2,6kahn1} However, such an understanding is much less advanced when a
lanthanide ion is involved. Magnetic investigations concerning molecular materials containing
lanthanide and transition metal ions have been overlooked until recently due to the weak
interactions present and the complications added by the orbital contribution of lanthanide
ions.~\cite{6dante} The case of Gd with a $^{8}$S$_{7/2}$ single-ion ground state and a quenched
orbital contribution is special; most of the interest has been focused on the Gd--Cu combination
because of the very frequently found ferromagnetic character of the Gd--Cu
interaction.~\cite{6gatt1,6gatt2,6gatt3,6bin1,6bin2,6bin3,6bin4,6theo1,6rfm,6theo2} Coupling those
blocks in a 3D lattice would produce molecular magnets and possibly ferromagnets.

In an extended polynuclear complex, namely Gd$_{2}$(ox)[Cu(pba)]$_{3}$\-[Cu\-(H$_{2}$O)$_{5}]\cdot
20$H$_{2}$O [hereafter abbreviated with Gd$_{2}$Cu$_{3}$(pba)] with
pba~$=$~1,3-pro\-pa\-ne\-diyl(oxa\-ma\-to), not only the ferromagnetic coupling of the Gd--Cu
interaction was recently reported, but also the onset of a long-range magnetic
order.~\cite{6lr,6emma} The crystal structure of this compound consists of layers of double-sheet
polymers separated by water molecules. A puckered arrangement of Gd[Cu(pba)] units forms a
two-dimensional honeycomb pattern connected by oxalato groups. Discrete [Cu(H$_{2}$O)$_{5}]^{2+}$
entities are interspersed in the gap between the layers. From specific heat measurements, the onset
of low-dimensional short-range order was found at temperatures around 1.5~K. By further lowering
the temperature, the phase transition to three-dimensional ferromagnetic long-range ordering was
observed at $T_{C}=1.05$~K. This compound represents the first molecular-based ferromagnet
involving lanthanide ions.

Although long-range ferromagnetic order has thus been found in Gd$_{2}$Cu$_{3}$(pba), this material
is isotropic due to the fact that both the Gd ions and the Cu ions are magnetically isotropic. A
successful method to increase the intrinsic anisotropy is to synthesize intermetallic lanthanide
(L) and transition metal (M) compounds, where the L has an orbital contribution, that introduces
the anisotropy that the transition metal is lacking. Such a strategy has of course been applied in
research on metallic ferromagnets, such as Nd$_{2}$Fe$_{14}$B. The strong orbit coupling within L,
on one side, and the intense L--M exchange coupling has the net effect of polarizing the two
sublattices in a direction, thus creating a strong uniaxial magnet, albeit, in a restricted
temperature region.~\cite{6inter} The same method in trying to increase the uniaxial anisotropy has
been applied in a series of molecular compounds based on L and M ions, by choosing the L that
induces such anisotropy (first condition), while expecting that the L--M interaction remains
ferromagnetic and of similar intensity to that in the Gd$_{2}$Cu$_{3}$(pba) compound (second
condition). The L anisotropy of the ground state depends on the crystal field acting on the orbital
moment.

In what follows, we shall describe the magnetic and thermal properties of a series of isostructural
complexes based on lanthanide and transition metal ions. The general chemical formula of the
compounds here studied is L$_{2}$[M(opba)]$_{3}\cdot x$DMSO$\cdot y$H$_{2}$O (hereafter abbreviated
with L$_{2}$M$_{3}$) where L~=~La, Gd, Tb, Dy, Ho and M~=~Cu, Zn, and opba stands for {\it
ortho}-phenylenebis\-(oxa\-ma\-to), while DMSO stands for dimethylsulfoxide. In view of previous
work (Ref.~\cite{6myrtil}) proving that the L--Cu exchange for light L substitutions is
antiferromagnetic, thus not fulfilling the second necessary condition above mentioned, we have
chosen to explore only the heavy L substitutions. Out of these, the most promising are the Tb and
Dy substitutions, where there are already evidences of ferromagnetic coupling with Cu from
susceptibility measurements, while no hint of such was found for Er, Tm or Yb.~\cite{6myrtil}
Recently, we have also reported, by means of specific heat experiments, the onset of long-range
magnetically ordered states in the Tb and Dy substituted compounds.~\cite{6icmm} In this article we
combine the thermal properties of L$_{2}$M$_{3}$ together with low-temperature susceptibility
measurements, and explain the obtained results by means of theoretical calculations.

Another interesting property of this series of isostructural complexes is that they exhibit a
spin-ladder structure. Spin-ladders are low-dimensional magnetic quantum systems that consist of
two or more strongly magnetically coupled chains of spins and are thus intermediate between one and
two dimensional magnetic systems. The magnetic properties of such systems have received a renewed
experimental and theoretical interest since the discovery of high-$T_{C}$ superconductivity in
ladder structures provided by some cuprates like SrCu$_{2}$O$_{3}$,~\cite{6ld1} or in vanadyl
pyrophosphate~\cite{6ld2} or in LaCuO$_{2.5}$.~\cite{6ld3} The research on novel molecular
complexes with ladder geometries is a very important task for a better understanding of the physics
behind these complex systems.~\cite{6ld4}

\section{Structural and experimental details}
\label{5struct}

Information on the structure of the L$_{2}$M$_{3}$ series of compounds has been provided in detail
in Refs.~\cite{6str,6str2}. Here, we summarize the most important information in view of the
analysis of the magnetic properties given below.

The compounds crystallize very poorly and, consequently, a complete X-ray diffraction analysis
could not be performed to determine the structure. Instead, the wide-angle X-ray scattering (WAXS)
technique has been used to obtain structural information. Even though the WAXS technique cannot
give the exact structure as compared to a full X-ray diffraction study, it has been successfully
applied in several cases, for instance to structurally characterize inorganic
polymers.~\cite{6myr3} Furthermore, a complete structure determination by X-ray diffraction has
been carried out on a related and very similar spin-ladder compound, namely,
Tm$_{2}$[Cu(opba)]$_{3}\cdot x$DMF$\cdot y$H$_{2}$O ($x\approx10~;~y\approx4$). This crystal
structure has therefore been utilized as a starting point to interpret the WAXS data for the
L$_{2}$M$_{3}$ compounds.

The analyses strongly suggested that the structure of L$_{2}$M$_{3}$ is likewise based on a
discrete, infinite ladder-like motif, as shown in Fig.~1. The sidepieces of the ladder consist of
L$_{2}$[M(opba)]$_{3}$ units developing along the $b$ direction with an alternation of L(III) and
M(II) ions. Rungs provided by the precursor [M(opba)]$^{2-}$, located between two L ions, connect
the sidepieces. The L--M distance across an oxamato bridge is $\approx5$~\AA, while the distance
between two L ions across a M(opba) group is $\approx11$~\AA. The shortest distance between two L
ions ($\approx10$~\AA) involves non-connected L ions belonging to two sidepieces of neighboring
ladders.

Each L ion is surrounded by seven oxygen ions, six of which belong to three opba ligands and one to
a water molecule. The L coordination polyhedrons may be described as capped trigonal antiprisms.
All the compounds are highly solvated with DMSO and H$_{2}$O molecules, and the exact number of
noncoordinated solvent molecules is not known accurately. Some of these molecules are easily
removed. Therefore, there is some uncertainty in the molecular weight of the compounds and
consequently in the absolute value of the molar susceptibility and specific heat.

The experiments done in the course of this work consisted primarily in measurements of three
quantities: magnetic susceptibility, magnetic moment and specific heat. Magnetic moment and
susceptibility data down to 1.7~K were obtained with a commercial SQUID-based magnetometer with an
ac-option. The ac-susceptibility of Gd$_{2}$Cu$_{3}$ in the temperature range 0.1~K~$<T<3~$K was
measured in Zaragoza with a mutual inductance technique in a dilution refrigerator. The excitation
amplitude was 10~mOe and the frequency $f=160$~Hz. The signal was measured by means of a
low-impedance ac-bridge, in which a SQUID was employed as a null detector.~\cite{6squid} The
low-temperature susceptibilities down to 5~mK of Tb$_{2}$Cu$_{3}$, Dy$_{2}$Cu$_{3}$ and
Ho$_{2}$Cu$_{3}$ were measured in Leiden with an ac-susceptometer directly mounted inside the
mixing chamber of a dilution refrigerator. The frequency of the experiments was $f=520$~Hz. The
data in the overlap region between 1.70~K and 3~K were used to convert the low-temperature data
from arbitrary into absolute units. The specific heat measurements on Gd$_{2}$Cu$_{3}$,
Gd$_{2}$Zn$_{3}$ and Dy$_{2}$Cu$_{3}$ in the range 0.2~K~$<T<5.7~$K were performed in Zaragoza by
using an adiabatic calorimeter cooled by adiabatic demagnetization, using the heat-pulse technique
and Ge thermometry. The error on the specific heat has been estimated to be less than
5\%.~\cite{phtr} The specific heats of Tb$_{2}$Cu$_{3}$, Ho$_{2}$Cu$_{3}$ and Ho$_{2}$Zn$_{3}$ were
measured down to 0.1 K in Leiden using a thermal relaxation technique. The calorimeter was mounted
in a dilution refrigerator and connected to a cold sink through a calibrated heat link. All data
were collected on powdered samples of the compounds.

\section{Magnetic properties above 2~K}
\label{5susce2}

Magnetic properties of systems containing lanthanide ions, such as L$_{2}$M$_{3}$, are
significantly influenced by the interaction between a lanthanide ion and the surrounding ions in
its direct environment. As a result, when a free lanthanide ion is placed in a crystal, its $2J+1$
fold degeneracy is partially lifted through electrostatic interaction between its $f$-electrons and
the charges of the surrounding ions. The multiplet is split into a number of states, which can
appropriately be termed the crystal field (CF) states. Magnetic susceptibility measurements may
offer valuable informations on the CF splittings. In this section we shall first present
ac-susceptibility data above 2~K of La$_{2}$Cu$_{3}$, Gd$_{2}$Cu$_{3}$ and Gd$_{2}$Zn$_{3}$ for
which the CF effects can be neglected. Thereafter we shall briefly review and re-analyze in terms
of the CF formalism the dc-susceptibility measurements on L$_{2}$M$_{3}$, with L~=~Tb, Dy, Ho and
M~=~Cu, Zn, which already appeared in Ref.~\cite{6myrtil}.

The ac-susceptibility of Gd$_{2}$Cu$_{3}$ was measured with a frequency of 90~Hz and an amplitude
of 4~Oe of the exciting field. The dc-measurements of Ref.~\cite{6myrtil} were performed in the
temperature range 2--300~K, with a magnetic field of $10^{3}$~Oe. Diamagnetic corrections of the
constituent atoms were estimated from Pascal's constants as $-291\times 10^{-6}$~emu/mol. Due to
the above mentioned uncertainty in the molecular weights, it was assumed that for each compound the
maximum expected value for the Curie constant was reached at 300~K, and the experimental data were
rescaled accordingly when necessary.

Hereafter we use the notation $J$, $J^{\prime}$ and $J^{\prime\prime}$ for the exchange constants
of the L--Cu, Cu--Cu and L--L interactions, respectively (negative values stand for
antiferromagnetic interactions).

\subsection{La$_{2}$Cu$_{3}$, Gd$_{2}$Cu$_{3}$ and Gd$_{2}$Zn$_{3}$}
\label{5xlacu}

The magnetism of La$_{2}$Cu$_{3}$, where the lanthanide ion is diamagnetic, allows an estimate of
the underlying Cu--Cu interaction. The single-ion magnetic properties of Cu(II) are fairly
straightforward. Spin-orbit coupling causes the $g$ values of the lowest Kramers doublet ($S=1/2$)
to lie in the range 2.0 to 2.3. Figure~2 shows the data of La$_{2}$Cu$_{3}$, which can be fitted to
a Curie-Weiss law with $\theta=-0.2$~K. This suggests that, even though the distances in the ladder
between next-nearest copper ions are as long as $\approx10$~\AA, they still interact with each
other with a weak antiferromagnetic superexchange coupling. Taking $S=1/2$, $\theta=-0.2$~K and
$z=4$ for the number of nearest neighbors for each copper ion, the mean-field expression for the
Curie-Weiss temperature~\cite{mfcurie} provides the value of $J^{\prime}/k_{{\rm B}}=-0.2$~K for
this Cu--Cu interaction.

Gadolinium(III) has a $^{8}$S$_{7/2}$ ground state; the orbital contribution is almost entirely
quenched and very isotropic $g$ values close to the free electron value are found. For temperatures
above 2~K, the in-phase ac-susceptibility of Gd$_{2}$Cu$_{3}$ follows a Curie-Weiss law with
$\theta=2.3$~K and $C=16.4$~emu~K~mol$^{-1}$, indicating the ferromagnetic nature of the compound
(Fig.~2). In the limit $T\gg\theta$, the paramagnetic susceptibility can be described by

\begin{equation}\label{5e3}
\chi\simeq\frac{C}{T}=\frac{2~C_{\mathrm{Gd}}+3~C_{\mathrm{Cu}}}{T},
\end{equation}

where $C_{\mathrm{Gd}}=N\mu_{eff}^{2}({\rm Gd})/3k_{{\rm B}}$ is the Curie constant of the two
gadolinium ions and $C_{\mathrm{Cu}}=N\mu_{eff}^{2}({\rm Cu})/3k_{{\rm B}}$ is the Curie constant
of the three copper ions. Taking $S_{\mathrm{Gd}}=7/2$, $S_{\mathrm{Cu}}=1/2$ and
$g_{\mathrm{Gd}}=g_{\mathrm{Cu}}=2.00$, one obtains

\begin{displaymath}
C=2~C_{\mathrm{Gd}}+3~C_{\mathrm{Cu}}=16.8~\mathrm{emu~K~mol}^{-1}
\end{displaymath}

which is in satisfactory agreement with the experimental value ($16.4$~emu K mol$^{-1}$). The same
mean-field analysis used above gives the value of $J/k_{{\rm B}}=0.5$~K for the Gd--Cu exchange
coupling. The origin of the ferromagnetic nature of the Gd--Cu interaction is further discussed in
Section~\ref{5disgdcu}.

When the copper in Gd$_{2}$Cu$_{3}$ is substituted by diamagnetic Zn, then the susceptibility of
Gd$_{2}$Zn$_{3}$ could be expected to follow the Curie law for two isolated Gd ions. Nevertheless,
Fig.~2 shows that the Curie behavior (with the expected value of $C=15.7$~emu K mol$^{-1}$) is no
longer maintained below 10~K, since a slight decrease of the $\chi T$ product from 15.7 down to
14.8 emu K mol$^{-1}$ at 2~K is observed. The data are better described by a Curie-Weiss law with
$\theta=-0.1$~K. Such a deviation may be due to a weak, antiferromagnetic Gd--Gd interaction of
dipolar origin. Since the Gd moment is large, dipole-dipole interactions could show up in the
susceptibility at temperatures as high as 10~K, given the nearest-neighbor Gd--Gd distance of
10~\AA. Of course it may also be attributed to a weak antiferromagnetic superexchange coupling
acting between the Gd ions. The mean-field analysis provides the value of $J^{\prime\prime}/k_{{\rm
B}}=-6\times10^{-3}$~K for this interaction, on basis of the obtained $\theta$-value.

\subsection{Tb$_{2}$Zn$_{3}$, Dy$_{2}$Zn$_{3}$ and
Ho$_{2}$Zn$_{3}$} \label{5xvarie}

As mentioned, for a lanthanide ion with a non-zero orbital moment, such as Tb, Dy or Ho, the effect
of the crystal field on the magnetic levels has to be taken into account. In order to study the CF
splittings for L$_{2}$M$_{3}$, we consider the susceptibility measurements for the M~=~Zn compounds
to avoid (or reduce) the complication of magnetic exchange interactions. To simplify the analysis
we consider the lanthanide ions in a cubic geometry, so that the number of crystal field parameters
is minimized. As a further simplification, we consider an octahedral coordination for each
lanthanide ion instead of the seven coordination found for Tm$_{2}$[Cu(opba)]$_{3}\cdot x$DMF$\cdot
y$H$_{2}$O ($x\approx10~;~y\approx4$), which crystallizes in orthorhombic
geometry.~\cite{6str,6str2} Although these simplifications may seem somewhat insatisfactory, the
lack of a precise knowledge of the actual coordination parameters (distances, angles) for the
various L ions in L$_{2}$M$_{3}$ does actually not justify more sophisticated treatments. Within
this approximation, the prediction of the type and the relative order of the CF split levels is
greatly simplified. We use the valuable paper of Lea, Leask and Wolf~\cite{6llw} (hereafter
referred to as LLW) which predicts both order and type of level in cubic symmetries for each value
of the angular momentum and for all ratios of the crystal field parameters.

The crystal field acting on the lanthanide ion in octahedral geometry can be represented by the
Hamiltonian

\begin{equation}\label{cfham}
{\mathcal
H}_{CF}=B_{4}(O_{4}^{0}+5O_{4}^{4})+B_{6}(O_{6}^{0}-21O_{6}^{4}),
\end{equation}

where the $O_{n}^{m}$ terms are angular momentum operators. The parameter $B_{n}$ is related to the
strength of the crystal field components

\begin{displaymath}
B_{n}=a_{n}A_{n}^{0}\langle r^{n}\rangle,
\end{displaymath}

where $a_{n}$ is the operator equivalent factor, $\alpha$, $\beta$, $\gamma$ for $n=2,4,6$,
respectively. $A_{n}^{0}$ are parameters relating the $n$th degree potential on the ion from the
ionic charges of the lattice, and $\langle r^{n}\rangle$ are the expectation values for the
distances of the $4f$ electrons. The product $A_{n}^{0}\langle r^{n}\rangle$ is determined to the
first order by geometric factors and in principle can be calculated.

LLW have determined the eigenfunction and eigenvalue solutions for the Hamiltonian equation for
applicable values of the angular momentum and the full range of $B_{4}/B_{6}$ ratios. They have
constructed diagrams for the eigenvalues of the crystal field levels of each angular momentum in
terms of a quantity $x$, defined as

\begin{displaymath}
\frac{x}{1-|x|}=\frac{F(4)B_{4}}{F(6)B_{6}}=\frac{F(4)\beta
A_{4}^{0}\langle r^{4}\rangle}{F(6)\gamma A_{6}^{0}\langle
r^{6}\rangle}=\frac{b_{4}}{b_{6}},
\end{displaymath}

where $-1<x<1$, and $F(4)$ and $F(6)$ are multiplicative factors. Then, the susceptibility can be
easily calculated for a given angular momentum and a gyromagnetic ratio. In the remainder of this
section, we discuss the susceptibility data~\cite{6myrtil} of L$_{2}$Zn$_{3}$ (with L~=~Tb, Dy, Ho)
compounds and we explain them in terms of the LLW scheme. Only the temperature region below 40~K
will be considered, since only the contribution of the lowest lying levels needs to be taken into
account. These approximations are allowed in view of the large size of the level splittings
compared to the low-temperature region of interest for our experiments (mainly below 10~K). This is
confirmed by the specific heat analyses given below, where we shall show that only few CF levels
are actually involved in the magnetic ordering processes.

Let us start with Tb$_{2}$Zn$_{3}$. Compounds of non-Kramers lanthanide ions, such as Tb(III),
often have a singlet electronic ground state separated by an energy $\Delta$ from an excited
singlet state. In a cubic field, the $^{7}F_{6}$ state of Tb is split into two singlets
($\Gamma_{1}$ and $\Gamma_{2}$), one non-magnetic doublet ($\Gamma_{3}$) and three triplets
[$\Gamma_{4}$, $\Gamma_{5}^{(1)}$ and $\Gamma_{5}^{(2)}$].

The measured susceptibility of Tb$_{2}$Zn$_{3}$ is shown in Fig.~3. The data at low temperatures
are well described by two singlets split by $\Delta\simeq 0.2$~K. In the LLW scheme for octahedral
coordination, such a situation corresponds to $|x|\approx 0.45$ where the ground state can be
either $\Gamma_{1}$ or $\Gamma_{2}$ (see Ref.~\cite{6llw}). The assignment will be confirmed by the
specific heat analysis presented below. For this simple case of two separated singlets, to improve
the quality of the fit, we used ultimately a model in which a weak antiferromagnetic exchange
interaction acting between the terbium ions has been taking into account. The fit provides the
value of $J^{\prime\prime}/k_{{\rm B}}=-1$~K for this interaction. The model used is explained in
detail in the Appendix and employed in Section~\ref{5totdis} to explain the very-low-temperature
magnetic and thermal properties of Tb$_{2}$Cu$_{3}$.

For the case of dysprosium(III), the crystal field splits the $^{6}H_{15/2}$ ground state into two
Kramers doublets ($\Gamma_{6}$ and $\Gamma_{7}$) and three quartets [$\Gamma_{8}^{(1)}$,
$\Gamma_{8}^{(2)}$ and $\Gamma_{8}^{(3)}$] (see Ref.~\cite{6llw}). In octahedral coordination with
$|x|\approx 0.45$, as estimated for the previous case of Tb, one anticipates a doublet as the
ground state ($\Gamma_{6}$ or $\Gamma_{7}$) with the higher state being the other doublet
($\Gamma_{7}$ or $\Gamma_{6}$, respectively). The fit to the Dy$_{2}$Zn$_{3}$ data (Fig.~3) is in
agreement with a low-lying doublet being either $\Gamma_{6}$ or $\Gamma_{7}$ separated by
$\Delta=13.7$~K from $\Gamma_{7}$ or $\Gamma_{6}$, respectively. Also this assignment will be
confirmed by the analysis of the specific heat (see below).

Holmium(III) is a non-Kramers ion, with a $^{5}I_{8}$ ground state. The crystal field splits the
ground state into one singlet ($\Gamma_{1}$), two non-magnetic doublets [$\Gamma_{3}^{(1)}$ and
$\Gamma_{3}^{(2)}$] and four triplets [$\Gamma_{4}^{(1)}$, $\Gamma_{4}^{(2)}$, $\Gamma_{5}^{(1)}$
and $\Gamma_{5}^{(2)}$]. The LLW diagram (see Ref.~\cite{6llw}) is complicated, with many level
crossings, which make simple choices difficult. For octahedral coordination with $|x|\approx 0.45$,
as in the Tb and Dy cases, one expects a $\Gamma_{3}^{(2)}$ ground state with $\Gamma_{4}^{(2)}$
the first excited state within a distance of not more than $\sim 20$~K. The fit to the
Ho$_{2}$Zn$_{3}$ data (Fig.~3) is in agreement with this, yielding the value of $\Delta=8.3$~K for
the $\Gamma_{3}^{(2)}$--$\Gamma_{4}^{(2)}$ separation.

To summarize, the susceptibility data presented here for the Tb, Dy and Ho containing compounds, as
well as the specific heat analyses presented below, can be reasonably accounted for by assuming
cubic symmetry and octahedral coordination for the lanthanide ions. The consistency of the analyses
is derived from the fact that the factor $x$ could be taken at approximately the same value
($\approx0.45$) for each compound. This is what one would expect for a series of isostructural
compounds in which the lanthanide ions most probably have the same local coordination.

\section{Very low-temperature properties}
\label{5susce1}

In the following sections, we present the thermal properties of the L$_{2}$M$_{3}$ compounds in the
temperature range 0.1 to 10~K, and their magnetic properties down to 5~mK. We first analyze the
lattice contributions to the measured heat capacities, and then describe the results inferred from
susceptibility, magnetization and magnetic specific heat for each compound of the series.

\subsection{Phonons contribution} \label{5cm}

Figure~4 shows the collected zero-field specific heat data of the L$_{2}$M$_{3}$ compounds as a
function of temperature. As a first step, we have estimated the lattice contribution for each
compound by fitting the high-temperature data to a sum of a Debye lattice contribution
$(C_{l}/R=\beta~T^{3})$ and a high-temperature limiting tail ($\propto T^{-2}$) of the magnetic
anomaly. The results for the $\beta$ parameter are given in Table~I.

The WAXS analysis has shown that the L$_{2}$M$_{3}$ compounds have similar structures (see
Section~\ref{5struct} and references therein). Consequently, one expects also a similar lattice
contribution for each of them. Indeed, the estimated $\beta$ values for five of the compounds
studied fall into a fairly narrow range between 1.1$\times10^{-2}$ and 3.2$\times10^{-2}$~K$^{-3}$.
Since the coefficient $\beta$ depends on the third power of the Debye temperature, such a limited
variation appears quite acceptable. The only exception is Dy$_{2}$Cu$_{3}$ for which
$\beta=8.8\times10^{-2}$~K$^{-3}$, a value much higher than the others. The reason is presently not
known. It is important to note that, however, the magnetic specific heat for Dy$_{2}$Cu$_{3}$
obtained after subtracting the phonon specific heat, appears to yield a magnetic entropy that
agrees with the value expected for a doublet ground state.

The magnetic contributions to the specific heats of all the L$_{2}$M$_{3}$ compounds discussed in
the following sections were obtained by subtracting the above mentioned lattice contributions.

\subsection{Gd$_{2}$Cu$_{3}$ and Gd$_{2}$Zn$_{3}$}
\label{5xlowgdcu}

We have already seen (Section~\ref{5xlacu}) that for temperatures above 2~K, the in-phase
susceptibility of Gd$_{2}$Cu$_{3}$ follows a Curie-Weiss law with $\theta=2.3$~K, indicating the
ferromagnetic nature of the compound. The low-temperature in-phase susceptibility of
Gd$_{2}$Cu$_{3}$ is depicted in Fig.~5 as a function of the temperature. A sharp peak is observed
at $T_{C}=1.78$~K which is probably due to a transition to long-range magnetic order. A large value
of the susceptibility is found at the maximum (65.9~emu/mol). This behavior is typical for a
powdered sample of an isotropic ferromagnetic material in which demagnetization effects become
important. The theoretical value estimated for the susceptibility of a ferromagnetic sample of
Gd$_{2}$Cu$_{3}$ at $T_{C}$ is $\chi_{ext}^{\prime}(T_{C})=N^{-1}\approx (70\pm 20)
~\mathrm{emu/mol}$, where $N$ is the demagnetizing factor calculated taking into account an
ellipsoidal approximation for the geometry of the sample. Within the error this is equal to the
experimental result.

An interesting feature, that is discussed later (Section~\ref{5disgdcu}), is the in-phase
susceptibility variation with the temperature for $T<T_{C}$ (Fig.~5). Just below $T_{C}$, the
susceptibility goes through a minimum and further lowering the temperature reveals a rounded
anomaly for 0.1~K~$<T<1~$K.

Figure~6 shows the variation of the experimental magnetization $M$ versus the field $H$ for the
temperature $T=1.70$~K, i.e. just below the critical temperature ($T_{C}=1.78$~K). No hysteresis
effect is observed. Additional evidence for the ferromagnetic coupling is gained by comparison of
the magnetization measurements to the calculated behavior for the uncoupled case. If the magnetic
centers were all uncoupled, $M$ would vary according to

\begin{equation}\label{5e4}
M=2Ng_{\mathrm{Gd}}\mu_{{\rm
B}}S_{\mathrm{Gd}}B_{7/2}(\eta_{\mathrm{Gd}})+
3Ng_{\mathrm{Cu}}\mu_{{\rm B}
}S_{\mathrm{Cu}}B_{1/2}(\eta_{\mathrm{Cu}})
\end{equation}

where $B_{7/2}(\eta_{\mathrm{Gd}})$ and $B_{1/2}(\eta_{\mathrm{Cu}})$ are the Brillouin functions
for Gd and Cu ions, respectively. The saturation magnetization would be given by

\begin{displaymath}
M=(2g_{\mathrm{Gd}}S_{\mathrm{Gd}}+3g_{\mathrm{Cu}}S_{\mathrm{Cu}})N\mu_{{\rm
B} } =17N\mu_{{\rm B}}.
\end{displaymath}

The calculated magnetization for $T=1.70$~K as derived from Eq.~(\ref{5e4}) is displayed in Fig.~6
together with the experimental curve. It can be observed that the experimental data saturate at the
predicted value and they lie above the expression of Eq.~(\ref{5e4}) in the whole temperature
range. Such a behavior clearly indicates that the predominant interaction, i.e. the Gd--Cu
interaction through the oxamato bridge, is ferromagnetic.

Let us now discuss the magnetic specific heat data by presenting first the result for the
Gd$_{2}$Zn$_{3}$ compound (Fig.~7). A sharp increase appears when lowering the temperature
below~1~K. Due to incomplete achievement of the ordering process in the temperature range of our
experiments, an analysis of the experimental entropy is not possible.

Figure~8 shows the magnetic molar specific heat $C_{m}/R$ versus the temperature $T$ for
Gd$_{2}$Cu$_{3}$. Two clear features can be observed: a distinct $\lambda$-peak at $T_{C}=(1.78\pm
0.02) $~K confirming the assignment of a long-range ordering process, and a rounded maximum around
0.8~K. We note that the rounded anomaly occurs in the same temperature range as that observed in
the susceptibility experiments (Fig.~5). Figure~8 shows also that the magnetic contribution at
$T>T_{C}$ is quite large and extends up to high temperatures, indicating the presence of important
short-range ordering effects most probably related to the low dimensionality (no contributions from
CF splittings are expected for Gd ions).

In order to calculate the entropy, we use the relation

\begin{equation}\label{2entropy}
S/R=\int_{0}^{\infty}(C_{{\rm m}}(T)/R\cdot T)~{\rm d}T,
\end{equation}

together with the experimental values of $C_{m}$ in the temperature range 0.2~K~$<T<5.7~$K. For the
extrapolation down to 0~K we assume a 3D ferromagnetic spin-wave type contribution ($\propto
T^{3/2}$) and on the high-temperature side a $T^{-2}$ dependence for $C_{m}$. The calculated
entropy gives a value of $S/R\approx 9~{\rm ln}~2$. This corresponds to the maximum expected value
($=3~{\rm ln}~2+2~{\rm ln}~8$) evidencing that both the Gd and Cu magnetic ions participate in the
ordering process. As already said, the relatively large values of the specific heat above $T_{C}$
may be associated with low-dimensional fluctuations within the ladders. This can be clearly seen by
plotting the entropy variation as a function of the temperature (inset of Fig.~8). The calculated
entropy variation above $T_{C}$ gives $(S_{\infty}-S_{C})\approx 2.7~{\rm R~ln}~2$, which is a
rather large fraction in comparison with theoretical values of three-dimensional models of
ferromagnets.~\cite{6mod} All this agrees with the low-dimensional ladder-type magnetic structure
of these materials.

\subsection{Tb$_{2}$Cu$_{3}$}
\label{5xlowtbcu}

The low-temperature susceptibility of Tb$_{2}$Cu$_{3}$ is depicted in Fig.~9. The abrupt change of
the in-phase susceptibility at 1~K is ascribed to a transition to a magnetically ordered state,
which is obvious also from the specific heat curve shown in Fig.~10. The maximum value is
93.6~emu/mol at 1.0~K. This value is almost of the same order as expected for a ferromagnetic
material in which demagnetization effects become important. In fact, an estimate of the
demagnetizing factor gives the value of $\approx 280$~emu/mol for the susceptibility of an
isotropic ferromagnetic Tb$_{2}$Cu$_{3}$ sample at the maximum, assuming a cylindrical
approximation of its shape. At the low temperature side of the anomaly, the susceptibility
decreases sharply, reaching a value of 5.0~emu/mol at 10~mK. The specific heat data are plotted in
Fig.~10 and indicate a transition temperature of $T_{C}=(0.81\pm 0.01)$~K. In the susceptibility
data we see that fluctuations in the ordering process show up as a peak in the out-of-phase
susceptibility $\chi^{\prime\prime}$ centered at 0.7~K (inset of Fig.~9), whereas the in-phase
susceptibility $\chi^{\prime}$ shows an inflection point (maximum in
$\partial\chi^{\prime}/\partial T$) at that same temperature. The $\chi^{\prime\prime}$ is in
arbitrary units because it is null above 0.8~K and, thus, it was not possible to scale to data
measured above 1.7~K with the commercial magnetometer. The observed behavior is not uncommon for a
low-dimensional system which undergoes a phase transition to long-range magnetic order, for
instance to an antiferromagnetic arrangement of the ferromagnetic ladders, at $T_{C}=0.81$~K as
deduced from the inflection point of $\chi^{\prime}$. The relatively large value of the
susceptibility at 1~K may suggest the presence of a dominating ferromagnetic interaction, probably
associated with the intraladder Tb--Cu interaction. We will return to this point in
Section~\ref{5totdis}. Moreover, the very low value reached by the in-phase susceptibility in the
limit of very low temperature, clearly indicates the large anisotropy of the terbium ions in
Tb$_{2}$Cu$_{3}$, as already pointed out in Section~\ref{5xvarie} for Tb$_{2}$Zn$_{3}$. The
anisotropy may also be responsible for the fact that the measured value of the $\chi^{\prime}$ at
$T_{C}$ is lower than the expected limit for an isotropic ferromagnetic sample. Since a powder is
measured, the value is a directional average.

The main feature of the magnetic specific heat of Tb$_{2}$Cu$_{3}$, depicted in Fig.~10, is the
$\lambda$-peak at $T_{C}=0.81$~K indicating the transition to a long-range ordered state. The bump
observed above $T_{C}$ and centered around 1.5~K is probably associated with low-dimensional
fluctuations within the ladders, since a contribution from excited CF-levels is not expected in
this temperature range (see Section~\ref{5xvarie}). The data at the low-temperature side show a
slight upward curvature. Below 0.4~K, the data obey the law $C_{m}T^{2}/R=0.05~{\rm K}^{2}$. This
contribution is probably coming from the hyperfine splitting of the magnetic levels of the Tb
nuclei. Subtracting the calculated hyperfine specific heat from the measured specific heat data, we
estimate the remaining entropy. In order to calculate the magnetic entropy of Tb$_{2}$Cu$_{3}$, we
carry out the integration in Eq.~(\ref{2entropy}) using the experimental values of $C_{m}/R$ in the
temperature range $0.1~{\rm K}<T<5$~K. After extrapolation down to 0~K with an exponential
function, the calculated entropy gives a value of $S/R\approx 5$~ln~2. Taking into account that Cu
has spin 1/2, which corresponds to $S/R=~$ln~2 per atom, the experimental entropy content indicates
that Tb has an effective spin 1/2, corresponding to a lowest lying doublet in the temperature
region of the magnetic ordering process. In Section~\ref{5xvarie} we have shown that for
Tb$_{2}$Zn$_{3}$ the lowest energy levels are two singlets with a separation of $\Delta=0.2$~K. For
Tb$_{2}$Cu$_{3}$ we expect a similar crystal environment and, as a consequence, a similar energy
scheme. Nevertheless, the separation $\Delta$ is relatively small in comparison with the thermal
energy in the region of the magnetic ordering process. We thus conclude that we can apply the
approximation of effective spin 1/2 for the Tb ion. In agreement with the pronounced
high-temperature specific heat tail observed in Fig.~10, only about 15\% of the entropy is left
below $T_{C}$. This is another indication of the large amount of short-range magnetic order that is
probably associated with the low-dimensionality of the ladder.

\subsection{Dy$_{2}$Cu$_{3}$}
\label{5xlowdycu}

Figure~11 shows the low-temperature ac-susceptibility of Dy$_{2}$Cu$_{3}$. As in the case of
Tb$_{2}$Cu$_{3}$, and as evidenced from the specific heat experiment presented in Fig.~12, clear
evidence of a transition to a magnetically ordered phase is found. The in-phase susceptibility has
a sharp peak centered at 0.8~K with a maximum value of 116.4~emu/mol. Below the peak, it decreases
very sharply and remains almost constant at 10.0~emu/mol down to 10~mK. As for Tb$_{2}$Cu$_{3}$,
the maximum of the experimental susceptibility is not far below the expected theoretical limiting
value ($\approx 280$~emu/mol) for an isotropic ferromagnetic sample. The out-of-phase
susceptibility also shows a sharp peak centered at 0.77~K and it is zero below 0.4~K and above 1~K
(inset of Fig.~11). At 0.77~K an inflection point occurs in the temperature dependence of the
$\chi^{\prime}$. The transition temperature deduced below from the specific heat is $T_{C}=(0.75\pm
0.01)$~K. Similar as for Tb$_{2}$Cu$_{3}$, the data point toward a type of antiferromagnetic
long-range ordering, with very small value for the antiferromagnetic coupling ($J_{AF}$) between
the ferromagnetic ladders. Since for an antiferromagnet the $\chi$ at $T_{C}$ is inversely
proportional to $J_{AF}$, when the latter becomes small enough, $\chi$ at $T_{C}$ can reach the
demagnetizing limit. It is then difficult to distinguish between ferro- or antiferromagnetic
coupling between the ferromagnetic chains. The peak in the $\chi^{\prime\prime}$ found in both
compounds may be due to a weak-ferromagnetic moment which arises when the antiferromagnetic
ordering is accompanied by some degree of spin canting. Again, as for Tb$_{2}$Cu$_{3}$, the very
low value of the in-phase susceptibility for $T\rightarrow 0$ indicates large anisotropy of the Dy
ions.

Figure~12 shows the magnetic specific heat, plotted as $C_{m}/R$ versus $T$, for the
Dy$_{2}$Cu$_{3}$ compound, where $C_{m}/R$ is the molar specific heat and $T$ is the temperature.
The prominent spike below 1~K is identified with the $\lambda$ anomaly indicating the onset of a
phase transition to a long-range ordered state. The peak is very sharp and allows an accurate
determination of the critical temperature as $T_{C}=(0.75\pm 0.01)$~K. The analysis of the magnetic
entropy content shows that only a very small portion (about 15\%) is obtained below $T_{C}$ and
that, as for Tb$_{2}$Cu$_{3}$, the lanthanide has an effective spin 1/2 in the ground state, once
more agreeing with the crystal field analysis (Section~\ref{5xvarie}).

\subsection{Ho$_{2}$Cu$_{3}$ and Ho$_{2}$Zn$_{3}$}
\label{5xlowhocu}

The low-temperature susceptibility of Ho$_{2}$Cu$_{3}$ is shown in Fig.~13. Also for this compound,
evidence of an antiferromagnetic long-range ordering is observed. The in-phase susceptibility has a
maximum of 27.5~emu/mol at 0.12~K (much lower than for Tb$_{2}$Cu$_{3}$ and Dy$_{2}$Cu$_{3}$). It
sharply decreases down to 8.9~emu/mol at 30~mK, and remains nearly constant by further lowering the
temperature down to 5~mK. The out-of-phase susceptibility shows a peak around 40--100~mK, in which
temperature range also an inflection point, centered at ($50\pm 10$)~mK, is seen in the in-phase
susceptibility.

The molar specific heat of Ho$_{2}$Cu$_{3}$ is depicted in Fig.~14 together with that of
Ho$_{2}$Zn$_{3}$ for comparison. The specific heats of Ho$_{2}$Cu$_{3}$ and Ho$_{2}$Zn$_{3}$ are
seen to overlap for $T>0.8$~K, showing a broad rounded anomaly with a maximum at 3.5~K (Fig.~14).
This is probably due to crystal field splitting of the holmium levels. In Section~\ref{5xvarie} we
have seen that the susceptibility of Ho$_{2}$Zn$_{3}$ down to 2~K may be explained in terms of a
doublet ($\Gamma_{3}^{(2)}$) ground state separated by 8.3~K from a triplet ($\Gamma_{4}^{(2)}$)
excited level. Taking into account a further splitting of these degenerate levels, the Schottky
contribution arising from this configuration can be easily calculated. A nice agreement with the
data is obtained for $\Gamma_{3}^{(2)}$ further split by 2~K, and separated by 11~K from
$\Gamma_{4}^{(2)}$, further split by 5~K each level (Fig.~14).

For Ho$_{2}$Zn$_{3}$, a sharp increase of the specific heat by lowering the temperature is observed
below 0.3~K. This contribution is apparently coming from the hyperfine splitting of the magnetic
levels of the Ho nuclei. In this temperature region, by fitting the data to $C_{m}/R=aT^{-b}$, we
obtain $a=0.14$ and $b=1.3$, and not the expected value of $b=2$, for a magnetic anomaly
high-temperature tail. The reason for that is probably related to problems of thermal contact
between spin system and the lattice, since we found the specific heat at low temperature to become
increasingly dependent on the measuring time we used in the experiments. In fact, both specific
heats of Ho$_{2}$Cu$_{3}$ and Ho$_{2}$Zn$_{3}$ have been measured with the relaxation technique. If
the measuring time is not long enough for the sample to achieve thermal equilibrium, between the
spin system and the phonon bath, part of the electronic spin contribution will simply be missing.
Also for Ho$_{2}$Cu$_{3}$ an upward curvature is observed by lowering the temperature below 0.7~K.
By comparison with the Ho$_{2}$Zn$_{3}$ results, we see that in the specific heat of
Ho$_{2}$Cu$_{3}$ there is an extra contribution besides the hyperfine splitting of the Ho nuclei.
This may readily be explained by a weak Ho--Cu coupling that, together with the dipole-dipole
interladder interaction, shows up at very low temperature. Consequently, only the onset of a phase
transition is observed. The low-temperature limit of our setup did not allow to determine the
ordering temperature that has to be below 0.1~K (judging from the $\chi$-data). In combination with
the susceptibility results (Fig.~13), we conclude that the ordering temperature $T_{C}$ for
Ho$_{2}$Cu$_{3}$ is between 0.04 and 0.10~K.

Due to the incomplete achievement of the magnetic ordering processes at the lowest temperature, the
analysis of the entropy contents for Ho$_{2}$Cu$_{3}$ and Ho$_{2}$Zn$_{3}$ is not possible.

\section{Comparison with ladder models}
\label{5dis}

\subsection{Gd$_{2}$Zn$_{3}$ and Gd$_{2}$Cu$_{3}$}
\label{5disgdcu}

For Gd$_{2}$Zn$_{3}$, the sharp increase of the specific heat below 1~K (Fig.~7) may reflect the
onset of a 3D ordering process, which takes place at a temperature lower than obtained in the
experiment. Even though the Gd$_{2}$Zn$_{3}$ compound is more poorly crystallized than the Cu
containing compounds,~\cite{6str2} its measured specific heat clearly underlines the relevance of
the Gd--Cu interaction, when compared with the results found for Gd$_{2}$Cu$_{3}$. The absence of
the copper at M sites reduces or inhibits the ordering temperature of the Gd sublattice below our
lowest experimental temperature. It is clear that the Cu ion in Gd$_{2}$Cu$_{3}$ is much more
effective than the Zn ion in transmitting magnetic exchange interaction along the
Gd-oxamato-Cu-oxamato-Gd superexchange pathway that gives rise to the net ferromagnetic coupling.

In order to explain the paramagnetic susceptibility of Gd$_{2}$Cu$_{3}$, let us consider the model
proposed by Georges et al.~\cite{6theo2} for ladder-type double chains which has been successfully
applied to describe the paramagnetic susceptibility of the
Gd$_{2}$(ox)[Cu(pba)]$_{3}$\-[Cu(H$_{2}$O)$_{5}]\cdot 20$H$_{2}$O compound. This model, which is
based on the standard transfer matrix method, takes into account the presence of copper quantum
spins and gadolinium classical spins, in a ladder-like arrangement similar to the L$_{2}$M$_{3}$
compounds, where each gadolinium ion interacts isotropically with two neighboring copper ions. The
approximation is the classical treatment of the gadolinium spin, which is allowed because of the
high spin value of the ion.

With the aid of this model, we have analyzed the experimental thermal dependence of the
susceptibility in the paramagnetic regime. The same Land\'e factor $g=2.00$, determined in the
high-temperature limit, has been attributed to all cations. A unique coupling constant $J$ has been
introduced for all the Gd--Cu isotropic interactions. The best fitting of the experimental
susceptibility results is obtained in the range 2.5~K~$<T<20~$K. Figure~5 shows the calculated
curve with $J/k_{{\rm B}}=0.74$~K. As expected, the positive $J$ value refers to a ferromagnetic
exchange coupling, in agreement with the positive paramagnetic Curie temperature $\theta=2.3$~K. It
is worthwhile to note that the $J$ value obtained here does not differ substantially from the
estimate of 0.5~K given in Section~\ref{5xlacu} deduced by simple mean field analysis. At
temperatures below 2.5~K, the three-dimensional ordering mechanism becomes apparent and,
consequently, the experimental data can no longer be described with the above (paramagnetic) model.

It is well established that an array of isotropic spins with dimensionality two or less will not
present long-range ordering.~\cite{6mod} The experimental specific heat data have shown the
presence of short-range order for temperatures far above $T_{C}$ already, and we have associated it
with extended magnetic correlations along the spin ladders. The observed long-range ordering at
$T_{C}=1.78$~K implies the existence of interactions between adjacent ladders. The structure of
Gd$_{2}$Cu$_{3}$ does not present any pathway between adjacent ladders that involve atomic contacts
suitable for the propagation of magnetic exchange. Consequently, it seems reasonable to assume that
the driving force for the magnetic long-range order is indeed the dipolar interaction.

Another interesting feature in Gd$_{2}$Cu$_{3}$ is the maximum observed at about 0.8~K, and thus
below $T_{C}=1.78$~K, in both the susceptibility (Fig.~5) and the specific heat (Fig.~8). To
explain such a feature, we may consider the presence of two types of interchain coupled systems in
Gd$_{2}$Cu$_{3}$, e.g. with slightly different packing of the ladders within the crystal structure.
Then one polytype has $T_{C}=1.78$~K and the other has a lower critical temperature of 0.8~K.
Unfortunately, due to the lack of a precise knowledge of the structure and to the fact that the
experiments have been performed on powdered samples, a detailed study of the ordering process could
not be carried out.

For the sake of completeness we wish to mention briefly the nature of the mechanism of the Gd--Cu
interaction. In preceding works,~\cite{6theo1,6rfm} the ferromagnetism of this interaction was
attributed to the coupling between the 4$f$--3$d$ ground configuration and the excited
configuration arising from the metal-metal charge transfer configurations. The latter is associated
with the $3d\rightarrow 5d$ process: an electron is transferred from the singly-occupied orbital
centered on copper toward an empty orbital centered on gadolinium. In such a mechanism, $J$ is
given by

\begin{equation}\label{5e6}
J=\sum_{i=1}^{5} [\beta_{5d-3d}^{2}\delta/(4U^{2}-\delta^{2})]_{i}
\end{equation}

where $\beta_{5d-3d}$ is the transfer integral of the $3d\rightarrow 5d$ process, $\delta$ is the
energy gap between $S=3$ and $S=4$ excited states arising from the $4f^{7}5d^{1}$ electron-transfer
configuration and $U$ is the energy cost of such a transfer. The summation applies to the five $5d$
gadolinium orbitals.

It is interesting to compare, qualitatively, the here obtained coupling constants $J$ of
Gd$_{2}$Cu$_{3}$ ($J/k_{{\rm B}}=0.74$~K) and of
Gd$_{2}$(ox)[Cu(pba)]$_{3}$[Cu(H$_{2}$O)$_{5}]\cdot 20$H$_{2}$O, which is a long-range ordered
ferromagnet for $T\le 1.05$~K~\cite{6lr,6emma} and has $J/k_{{\rm B}}=0.40$~K.~\cite{6theo2} The
difference may in fact be understood on the basis of the above mechanism. In the latter compound,
the Gd ions adopt a distorted capped square antiprism coordination of oxygen atoms; six out of a
total of nine oxygen atoms are provided by the three bridging Cu(pba) groups, two by the oxalato
ligand, and the last one by a water molecule.~\cite{6stru} In the former compound, due to the
absence of the oxalato ligand, the $5d$ orbitals of the gadolinium ions are distributed in a less
distorted environment. This fact, together with slightly shorter Gd--Cu distances for
Gd$_{2}$Cu$_{3}$ (see Refs.~\cite{6str2} and \cite{6stru} for comparison), gives higher values of
$\beta_{5d-3d}$ for Gd$_{2}$Cu$_{3}$. Consequently, according to Eq.~(\ref{5e6}), the coupling
constant $J$ should indeed be larger for Gd$_{2}$Cu$_{3}$ than for
Gd$_{2}$(ox)[Cu(pba)]$_{3}$[Cu(H$_{2}$O)$_{5}]\cdot 20$H$_{2}$O, as found.

\subsection{Tb$_{2}$Cu$_{3}$, Dy$_{2}$Cu$_{3}$ and
Ho$_{2}$Cu$_{3}$}\label{5totdis}

Our measurements show Tb$_{2}$Cu$_{3}$, Dy$_{2}$Cu$_{3}$ and Ho$_{2}$Cu$_{3}$ to be strongly
ani\-so\-tro\-pic systems with long-range order phase transitions below 1~K. As for
Gd$_{2}$Cu$_{3}$, the low-dimensional structure of the compounds implies that the driving force for
the long-range magnetic order is the dipolar interaction between the ferromagnetic chains. However,
for Tb$_{2}$Cu$_{3}$ and Dy$_{2}$Cu$_{3}$, evidence for an even larger amount of short-range order,
associated with fluctuations within the ladder structure, is found above $T_{C}$. From the specific
heat data, we have deduced that Tb$_{2}$Cu$_{3}$ and Dy$_{2}$Cu$_{3}$ may be described in terms of
an effective spin 1/2 at low temperatures. We note that an effective spin 1/2, arising from a
low-lying doublet, is often found for lanthanide compounds of dysprosium and terbium with low
magnetic ordering temperature.~\cite{6icmm5} At higher temperatures the behavior is strongly
modified by the presence of low-lying excited states.

To explain the magnetic and thermal properties of Tb$_{2}$Cu$_{3}$ and Dy$_{2}$Cu$_{3}$, a model
has been developed, which is based on the standard transfer matrix technique, using the Ising
approximation and assuming spin 1/2 for the L and the Cu ions. The details of the model can be
found in the Appendix. As revealed from the susceptibility measurements, we have to expect a
ferromagnetic coupling $J$ for the L--Cu interaction. Next-nearest neighbor interactions are also
taken into consideration. In fact, we have seen in Section~\ref{5xlacu} that a weak
antiferromagnetic interaction is acting between the copper ions in La$_{2}$Cu$_{3}$. Consequently,
we may assume a similar Cu--Cu interaction to be present in L$_{2}$Cu$_{3}$ for L~=~Tb and Dy. We
shall denote by $J^{\prime}$ the Cu--Cu interaction. Similarly, we have also taken into account a
weak antiferromagnetic coupling $J^{\prime\prime}$ acting between the L ions, as suggested by the
susceptibility data for Gd$_{2}$Zn$_{3}$ (Section~\ref{5xlacu}). According to our notation, a
positive coupling constant implies a ferromagnetic coupling. The presence of next-nearest neighbor
interactions in 1D molecular systems based on lanthanide ions is not new and several examples can
be found in the literature.~\cite{6dante1d} For the case of Tb$_{2}$Cu$_{3}$, the crystal field
energy separation $\Delta$ of the two lowest singlets of the non-Kramers Tb ion has also been
included in the calculation.

To improve the reliability of the fits for Tb$_{2}$Cu$_{3}$ and Dy$_{2}$Cu$_{3}$, we independently
use the model in La$_{2}$Cu$_{3}$ to get the estimate of $J^{\prime}$, and in Tb$_{2}$Zn$_{3}$ and
Dy$_{2}$Zn$_{3}$ to get the estimates of $J^{\prime\prime}$. We then fit the susceptibilities and
specific heats of Tb$_{2}$Cu$_{3}$ and Dy$_{2}$Cu$_{3}$ with $J^{\prime}$ and $J^{\prime\prime}$
fixed at the values obtained precedently with only the coupling $J$ and the effective gyromagnetic
constant $g$ as free parameters. The fits to the susceptibility and specific heat of
Tb$_{2}$Cu$_{3}$ are shown in Figs.~9 and 10, respectively. The fits provide the values of
$J/k_{{\rm B}}=5.6$~K, $J^{\prime}/k_{{\rm B}}=-1.0$~K, $J^{\prime\prime}/k_{{\rm B}}=-1.0$~K,
$\Delta=0.2$~K and an effective gyromagnetic constant of $g=18.0$ for the Tb ion. Similarly, the
results for the susceptibility and specific heat of Dy$_{2}$Cu$_{3}$ are shown in Figs.~11 and 12,
respectively. In this case, the fits provide the values of $J/k_{{\rm B}}=4.6$~K,
$J^{\prime}/k_{{\rm B}}=-1.0$~K, $J^{\prime\prime}/k_{{\rm B}}=-1.0$~K, and effective $g=19.6$. It
can be seen that the experimental behaviors above the 3D ordering temperatures are satisfactory
reproduced. Only the specific heat of Tb$_{2}$Cu$_{3}$ is not so well accounted for (Fig.~10).
Probably, this comes from the influence of low-lying excited levels, making the Hamiltonian of this
system less Ising-like than for Dy$_{2}$Cu$_{3}$. Moreover, the large $g$ values found for both
systems imply large magnetic moments which are moreover coupled ferromagnetically into chains, so
that dipolar interactions may be strong enough to contribute also above the 3D ordering
temperatures. Consequently, the model here presented may fail to give a detailed explanation in
this temperature region. Nevertheless, the following conclusions can be drawn. The compounds
Tb$_{2}$Cu$_{3}$ and Dy$_{2}$Cu$_{3}$ are very similar in behavior and are strongly anisotropic. We
note that large $g$ values of order 20, found for both Tb$_{2}$Cu$_{3}$ and Dy$_{2}$Cu$_{3}$, have
been also reported for other Dy and Tb based compounds.~\cite{6carlin} For both cases, the L--Cu
interaction is predominant and ferromagnetic. This interaction is slightly stronger for
Tb$_{2}$Cu$_{3}$ than for Dy$_{2}$Cu$_{3}$, in line with the analysis reported in
Ref.~\cite{6myrtil}. Consequently, also the 3D ordering temperature is higher for Tb$_{2}$Cu$_{3}$
($T_{C}=0.81$~K) than for Dy$_{2}$Cu$_{3}$ ($T_{C}=0.75$~K). For both compounds, the presence of
next-nearest neighbor interactions have to be taken into account. These interactions are much
weaker than the L--Cu interaction and are antiferromagnetic in nature. In conclusion, the results
here obtained suggest that Tb$_{2}$Cu$_{3}$ and Dy$_{2}$Cu$_{3}$ order ferromagnetically within the
ladders. This ordering process is accompanied by weak dipolar and probably antiferromagnetic
interladder interactions, that, together with strong crystal field effects, lower the
susceptibility below $T_{C}$.

For Ho$_{2}$Cu$_{3}$, we have already reported that the Ho--Cu interaction has to be very weak.
Indeed, crystal field effects and dipolar interactions are mainly responsible for the magnetic
properties at low temperature (Section~\ref{5xlowhocu}). Consequently, also the 3D ordering
temperature is the lowest one in comparison with Tb$_{2}$Cu$_{3}$ and Dy$_{2}$Cu$_{3}$.

The results obtained for the magnetic interaction parameters are summarized in Table~II for the Cu
containing compounds and in Table~III for the Zn containing compounds.

\section{Concluding remarks} \label{5discr}

In the previous sections we have focused attention on the physical properties of spin-ladder
molecular magnets containing lanthanide and transition metal ions. Together with the originality of
their crystal structure, the L$_{2}$M$_{3}$ compounds present interesting magnetic features such as
the onset of long-range orderings for L~=~Gd, Tb, Dy, Ho and M~=~Cu. To the best of our knowledge,
Gd$_{2}$Cu$_{3}$ is the lanthanide and transition metal ions based ferromagnet with the highest
long-range ordering critical temperature $T_{C}=(1.78\pm 0.02)$~K so far reported. Moreover, as
estimated from specific heat and susceptibility measurements, Tb$_{2}$Cu$_{3}$ has a magnetic
ordering temperature of $T_{C}=(0.81\pm 0.01)$~K, while Dy$_{2}$Cu$_{3}$ orders at $T_{C}=(0.75\pm
0.01)$~K, and Ho$_{2}$Cu$_{3}$ has a $T_{C}$ between 0.04 and 0.10~K. These molecular based magnets
are the first ones obtained with lanthanide ions other than gadolinium.

The very pronounced quasi-one-dimensionality of the magnetic structure implies that the 3D ordering
is driven by the dipolar interaction acting between ladders together with the intraladder
superexchange interaction. We have also reported that the intraladder L--Cu interaction is the
dominant one and it is ferromagnetic for L~=~Gd, Tb and Dy. The influence on the magnetic
properties of the weaker antiferromagnetic next-nearest neighbor L--L and Cu--Cu interactions have
also been analyzed. A remarkable point is the key role of the copper ions in these complex systems.
They transmit the magnetic exchange interaction between the L ions across the ladder. As a proof of
this, we have seen that, if Cu is substituted by the non-magnetic Zn, the L--L interaction is
inhibited and the critical temperature of the 3D ordering process is strongly diminished. Finally,
the effects on the susceptibility and specific heat of the crystal field splittings of the magnetic
energy levels of the lanthanide ions in L$_{2}$M$_{3}$ have been discussed and rationalized,
assuming the same symmetry and coordination for the lanthanide ion in each compound of the series.

\section*{APPENDIX: ISING MODEL FOR QUANTUM SPIN LADDER}
\label{5model}

The model here reported has been developed to explain the magnetic and thermal properties of the
Tb$_{2}$Cu$_{3}$ and Dy$_{2}$Cu$_{3}$ compounds.

As a first step, we assume only a coupling between the next-nearest copper (Cu) and lanthanide
(L~=~Tb, Dy) ions which is ferromagnetic (in the following notations, this means $J>0$).

For Tb and Dy, we assume an effective low-temperature spin $1/2$ and Ising coupling due to the
crystal field anisotropy of the lanthanide ion. For Cu, we assume spin $1/2$ as well. The effective
Hamiltonian is then:

\begin{equation}\label{appex1}
{\mathcal H}=-J\sum_{\langle i,j\rangle}(\sigma_{i}S_{j})
-\Bigl(\sum_{i=1}^{N_{\sigma}}g_{\sigma}\mu_{{\rm
B}}\sigma_{i}+\sum_{j=1}^{N_{S}}g_{S}\mu_{{\rm
B}}S_{j}\Bigr)H_{z},
\end{equation}

where $\sigma_{i}$ and $S_{j}$ are the projections along the $z$ axis of the spins of the copper
and lanthanide ions, respectively (Fig.~15). The Land\'e factor of the copper ions is assumed to be
$g_{\sigma}=2$, while the one of the lanthanide ions is a fitting parameter. The Zeeman terms in
Eq.~(\ref{appex1}) extend over all $N_{\sigma}$ copper and $N_{S}$ lanthanide ions.

We build the transfer matrix on the unit cell

\begin{widetext}

\begin{eqnarray*}
T(S_{1},S_{2};S_{1}^{\prime},S_{2}^{\prime})&=&
\sum_{\sigma_{1},\sigma_{2},\sigma_{3},\sigma_{4}}~{\rm exp}~\biggl\{\beta
J\biggl[\sigma_{1}(S_{1}+S_{1}^{\prime})
+\sigma_{2}(S_{2}+S_{2}^{\prime})+~\frac{1}{2}~\sigma_{3}(S_{1}+S_{2}^{\prime})
+\frac{1}{2}~\sigma_{4}(S_{1}+S_{2}^{\prime})\biggr]\biggr\}\\ &&\times~{\rm
exp}~\biggl\{\frac{1}{2}~\beta g_{S}\mu_{{\rm
B}}H_{z}(S_{1}+S_{1}^{\prime}+S_{2}+S_{2}^{\prime})+~\beta g_{\sigma}\mu_{{\rm
B}}H_{z}\biggl(\sigma_{1}+\sigma_{2}+\frac{1}{2}~\sigma_{3}+ \frac{1}{2}~\sigma_{4}\biggr)\biggr\}
\end{eqnarray*}

\end{widetext}

where $\beta=1/k_{{\rm B}}T$. The partition is expressed in terms of the $4\times 4$ transfer
matrix

\begin{eqnarray*}
Z(T,H_{z})&=&\sum_{S_{i}}
T(S_{1},S_{2};S_{1}^{\prime},S_{2}^{\prime})
T(S_{1}^{\prime},S_{2}^{\prime};S_{1}^{\prime\prime},S_{2}^{\prime\prime})...\\
&&...T(S_{1}^{(N)},S_{2}^{(N)};S_{1},S_{2})={\rm Tr}~(T^{N})
\end{eqnarray*}

Therefore, by defining

\begin{displaymath}
f(T,H_{z})\equiv\lim_{N\rightarrow\infty}\frac{1}{N}~{\rm ln}~Z
={\rm ln}~\lambda
\end{displaymath}

where $\lambda$ is the largest eigenvalue of $T$, we obtain numerically the magnetic susceptibility
and the molar specific heat by

\begin{displaymath}
\chi=k_{{\rm B}}T\frac{\partial^{2}f}{\partial H_{z}^{2}}\qquad ;
\qquad C=k_{{\rm
B}}\beta^{2}\frac{\partial^{2}f}{\partial\beta^{2}}.
\end{displaymath}

According to the susceptibility data of La$_{2}$Cu$_{3}$ (Fig.~2), we assume a slight
antiferromagnetic next-nearest neighbor coupling $J^{\prime}$ between the copper ions. Similarly,
according to the susceptibility data of Gd$_{2}$Zn$_{3}$ (Fig.~2), we expect an antiferromagnetic
next-nearest neighbor coupling $J^{\prime\prime}$ between the lanthanide ions. In both cases, the
data are compatible with couplings not exceeding 1~K. By adding these second neighbors couplings we
get the complete scheme depicted in Figure~15.

The transfer matrix is now

\begin{widetext}

\begin{eqnarray*}
T(S_{1},\sigma_{3},S_{2};S_{1}^{\prime},\sigma_{4},S_{2}^{\prime})&=
&\sum_{\sigma_{1},\sigma_{2}}~{\rm exp}~\biggl\{\beta J\biggl[\sigma_{1}(S_{1}+S_{1}^{\prime})
+\sigma_{2}(S_{2}+S_{2}^{\prime})+~\frac{1}{2}~\sigma_{3}(S_{1}+S_{2}^{\prime})
+\frac{1}{2}~\sigma_{4}(S_{1}+S_{2}^{\prime})\biggr]\biggr\}\\ &&\times~{\rm exp}~\biggl\{\beta
J^{\prime}(\sigma_{1}+\sigma_{2})(\sigma_{3}+\sigma_{4})+~\beta
J^{\prime\prime}\biggl[S_{1}S_{1}^{\prime}+S_{2}S_{2}^{\prime}+\frac{1}{2}~
(S_{1}S_{2}+S_{1}^{\prime} S_{2}^{\prime})\biggr]\bigg\}\\ &&\times~{\rm
exp}~\biggl\{\frac{1}{2}~\beta g_{S}\mu_{{\rm
B}}H_{z}(S_{1}+S_{1}^{\prime}+S_{2}+S_{2}^{\prime})+~\beta g_{\sigma}\mu_{{\rm
B}}H_{z}\biggl(\sigma_{1}+\sigma_{2}+\frac{1}{2}~\sigma_{3}+ \frac{1}{2}~\sigma_{4}\biggr)\biggr\}
\end{eqnarray*}

\end{widetext}

with $J>0$, $J^{\prime}<0$ and $J^{\prime\prime}<0$. This is now an $8\times 8$ matrix but the
process is the same.

For the terbium case, that is a non-Kramers ion, we allow the spin states to be non degenerate by
adding to the Hamiltonian a term $E(S_{i})$ such that $E(+\frac{1}{2})=\frac{\Delta}{2}$ and
$E(-\frac{1}{2})=-\frac{\Delta}{2}$, where $\Delta$ is the gap. The best fits for Tb$_{2}$Cu$_{3}$
are displayed in Figs.~9, 10 and for Dy$_{2}$Cu$_{3}$ in Figs.~11, 12.

\section*{ACKNOWLEDGMENTS}
We are indebted to F.L. Mettes, F. Luis, A. Morello, and D. Culebra for the low-temperature
experiments. This work was partially supported by Grant No. MAT99/1142 from CICYT.

\newpage

\begin{table}[h]
\label{5tab1}
\begin{tabular}{|c ||c | c| c| c| c| c|}
\hline &Gd$_{2}$Cu$_{3}$ &Tb$_{2}$Cu$_{3}$ &Dy$_{2}$Cu$_{3}$ &Ho$_{2}$Cu$_{3}$ &Gd$_{2}$Zn$_{3}$
&Ho$_{2}$Zn$_{3}$\\ \hline $\beta$ &3.2 &1.5 &8.8 &1.2 &1.1 &2.9
\\ \hline
\end{tabular}
\caption{Experimental lattice contributions $(C_{l}/R=\beta~T^{3})$ as estimated from the data in
Fig.~4. The values $\beta$ are expressed in $(\times10^{-2}~K^{-3})$.}
\end{table}

\begin{table}[ht]
\label{5tabtot2}
\begin{tabular}{c |c  c c c c}
&La$_{2}$Cu$_{3}$& Gd$_{2}$Cu$_{3}$& Tb$_{2}$Cu$_{3}$ &Dy$_{2}$Cu$_{3}$ &Ho$_{2}$Cu$_{3}$\\ \hline
&&&&&\\ $T_{C}$ (K) &&1.78(2) &0.81(1) &0.75(1) &$0.04\div 0.10$\\ $\theta$ (K)&$-0.2$
&$2.3$&$$&$$&$$\\ $J/k_{{\rm B}}$ (K) &&0.74&5.6&4.6&\\ $J^{\prime}/k_{{\rm B}}$
(K)&$-0.2$&&$-1.0$&$-1.0$&\\ $J^{\prime\prime}/k_{{\rm B}}$ (K) &&&$-1.0$&$-1.0$&\\ $\Delta$
(K)&&&0.2 (s--s)&&$\approx 11$ (d--t)\\ $g_{{\rm L}}$&&2&18.0&19.6&\\
Model&MF&Georges&App.&App.&Schottky-LLW\\
\end{tabular}
\caption{Experimental results obtained for the Cu containing compounds. $T_{C}$ is the long-range
magnetic ordering temperature; $\theta$ is obtained from the Curie-Weiss law
(Section~\ref{5susce2}); $J$, $J^{\prime}$ and $J^{\prime\prime}$ are the exchange constants for
the L--Cu, Cu--Cu and L--L interactions, respectively (negative values stand for antiferromagnetic
interactions); $\Delta$ is the separation between the ground state and the first excited state
(included also is the type of state: s=singlet; d=doublet; t=triplet); $g_{{\rm L}}$ is the Land\'e
factor for the L ion; the models used are MF (mean field) or LLW (Section~\ref{5xvarie}) or Georges
(Section~\ref{5disgdcu}) or App. (Appendix and Section~\ref{5totdis}) or Schottky
(Section~\ref{5xlowhocu}).}
\end{table}

\begin{table}[!ht]
\label{5tabtot1}
\begin{tabular}{c |c  c c c}
&Gd$_{2}$Zn$_{3}$ &Tb$_{2}$Zn$_{3}$ &Dy$_{2}$Zn$_{3}$ &Ho$_{2}$Zn$_{3}$\\ \hline &&&&\\ $\theta$
(K) &$-0.1$ &$$ &$$ &$$\\ $J^{\prime\prime}/k_{{\rm B}}$ (K)&$-6\times 10^{-3}$&$-1.0$&&\\ $\Delta$
(K)&&0.2 (s--s)&13.7 (d--d)& 8.3 (d--t)\\ Model &MF&LLW-App.&LLW&LLW\\
\end{tabular}
\caption{Experimental results obtained for the Zn containing compounds. The same notation of
Table~II is used.}
\end{table}

\newpage

LIST OF FIGURES

\begin{enumerate}

% fig1
\item{Relative dispositions of the ladders: (a) view of two neighboring ladders in the $bc$ plane;
(b) schematic view of the ladders projected in the $ac$ plane. The lanthanide ions are located on
the sides of the ladder and occupy each corner, while each transition metal ion is between two
lanthanide ions along the sides and in the rungs.}

% fig2
\item{TOP: Experimental molar susceptibility of La$_{2}$Cu$_{3}$. In the inset: the same data plotted as
$\chi T$ vs $T$. BOTTOM: Inverse of the in-phase susceptibilities of Gd$_{2}$Zn$_{3}$ ($\bullet$)
and Gd$_{2}$Cu$_{3}$ ($\circ$). In the inset: $\chi T$ vs $T$ for Gd$_{2}$Zn$_{3}$.}

% fig3
\item{TOP: Experimental inverse susceptibility for Tb$_{2}$Zn$_{3}$. The solid line is the calculated
inverse susceptibility taking into account $\Delta=0.2$~K and $J^{\prime\prime}/k_{{\rm B}}=-1$~K
for the Tb--Tb interaction. For explanations, see text. CENTER: Measured and calculated inverse
susceptibility for Dy$_{2}$Zn$_{3}$. BOTTOM: Measured and calculated inverse susceptibility for
Ho$_{2}$Zn$_{3}$.}

% fig4
\item{Experimental specific heats of L$_{2}$M$_{3}$.}

% fig5
\item{Experimental molar in-phase susceptibility of Gd$_{2}$Cu$_{3}$ versus temperature and
theoretical estimation for $J/k_{{\rm B}}=0.74$~K (see Section~\ref{5disgdcu} for explanations).}

% fig6
\item{Field dependence of the magnetization and hysteresis loop at $T=1.70$~K for Gd$_{2}$Cu$_{3}$.
The dotted line represents the Brillouin functions as calculated from Eq.~(\ref{5e4}).}

% fig7
\item{Magnetic specific heat vs temperature for Gd$_{2}$Zn$_{3}$. Solid curve is a guide to the
eye.}

% fig8
\item{Magnetic molar specific heat vs temperature for Gd$_{2}$Cu$_{3}$. Curve $a$ is the
ferromagnetic spin wave contribution $(\propto T^{3/2})$; curve $b$ is the high temperature limit
$(\propto T^{-2})$ of the magnetic anomaly tail. In the inset: temperature dependence of the
magnetic entropy for Gd$_{2}$Cu$_{3}$.}

% fig9
\item{Experimental low-temperature in-phase susceptibility vs temperature for Tb$_{2}$Cu$_{3}$
together with the calculated susceptibility (for explanations see Section~\ref{5totdis}). In the
inset: Experimental out-of-phase susceptibility.}

% fig10
\item{Magnetic molar specific heat of Tb$_{2}$Cu$_{3}$. The dotted line is the calculated hyperfine
contribution of the Tb ions. The solid line is the calculated specific heat due to the
low-dimensionality of the ladder (for explanations see Section~\ref{5totdis}).}

% fig11
\item{Experimental low-temperature in-phase susceptibility vs temperature for Dy$_{2}$Cu$_{3}$
together with the calculated susceptibility (for explanations see Section~\ref{5totdis}). In the
inset: Experimental out-of-phase susceptibility.}

% fig12
\item{Magnetic molar specific heat of Dy$_{2}$Cu$_{3}$. The solid line is the calculated specific
heat due to the low-dimensionality of the ladder (for explanations see Section~\ref{5totdis}).}

% fig13
\item{Experimental molar in-phase and out-of-phase (inset) susceptibilities vs temperature for
Ho$_{2}$Cu$_{3}$. Solid curves are guides to the eye.}

% fig14
\item{Magnetic molar specific heat of Ho$_{2}$Cu$_{3}$ $(\bullet)$ and Ho$_{2}$Zn$_{3}$ $(\circ)$.
The solid line is the calculated Schottky contribution due to the splitting of the
$\Gamma_{3}^{(2)}$ and $\Gamma_{4}^{(2)}$ levels of the holmium ions as shown in the Figure. In the
inset the same data in log-log scale.}

% fig15
\item{Ladder structure considered by the model discussed in the Appendix.}
\end{enumerate}

\clearpage
\includegraphics[width=17cm]{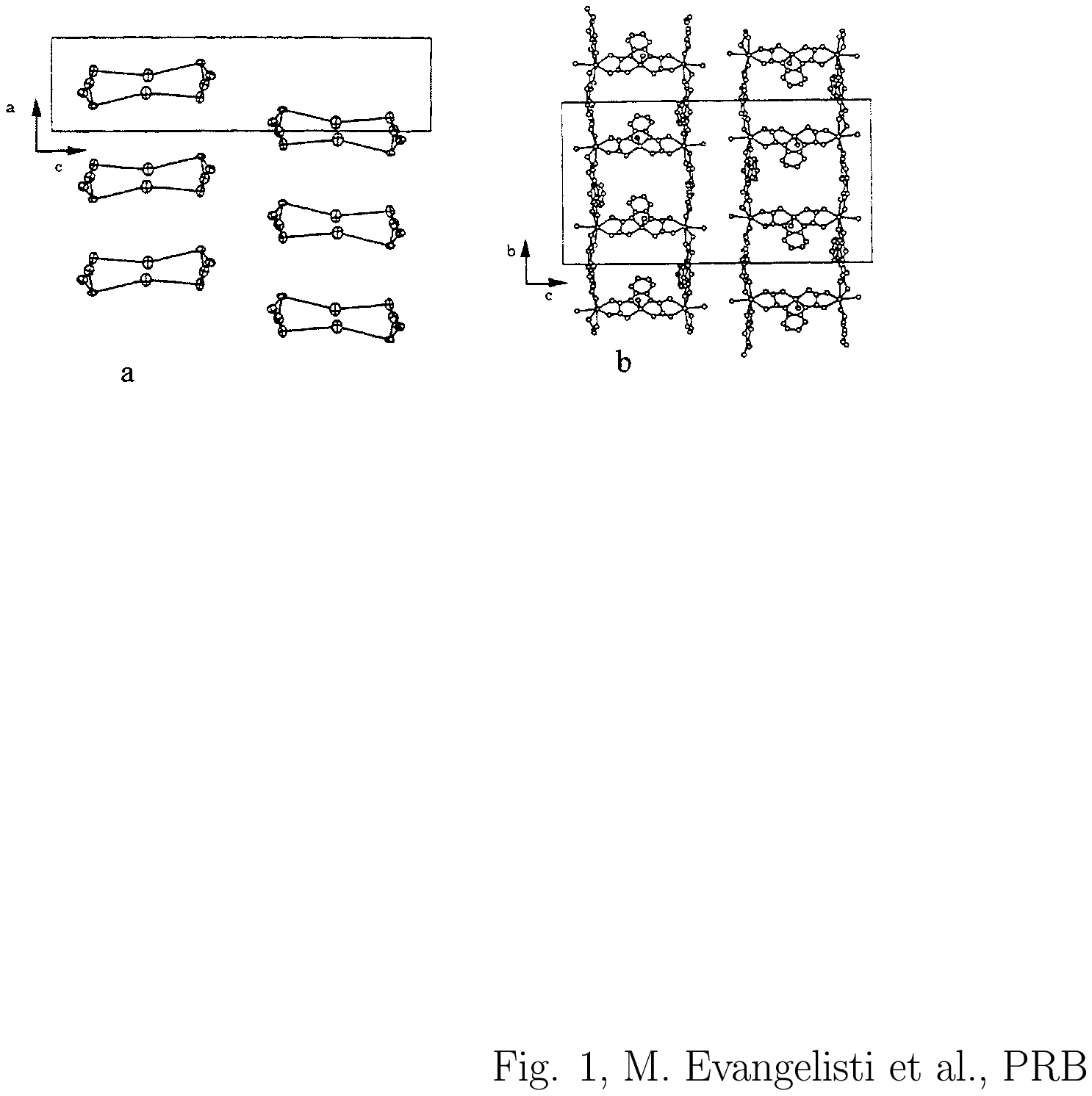}
\clearpage
\includegraphics[width=17cm]{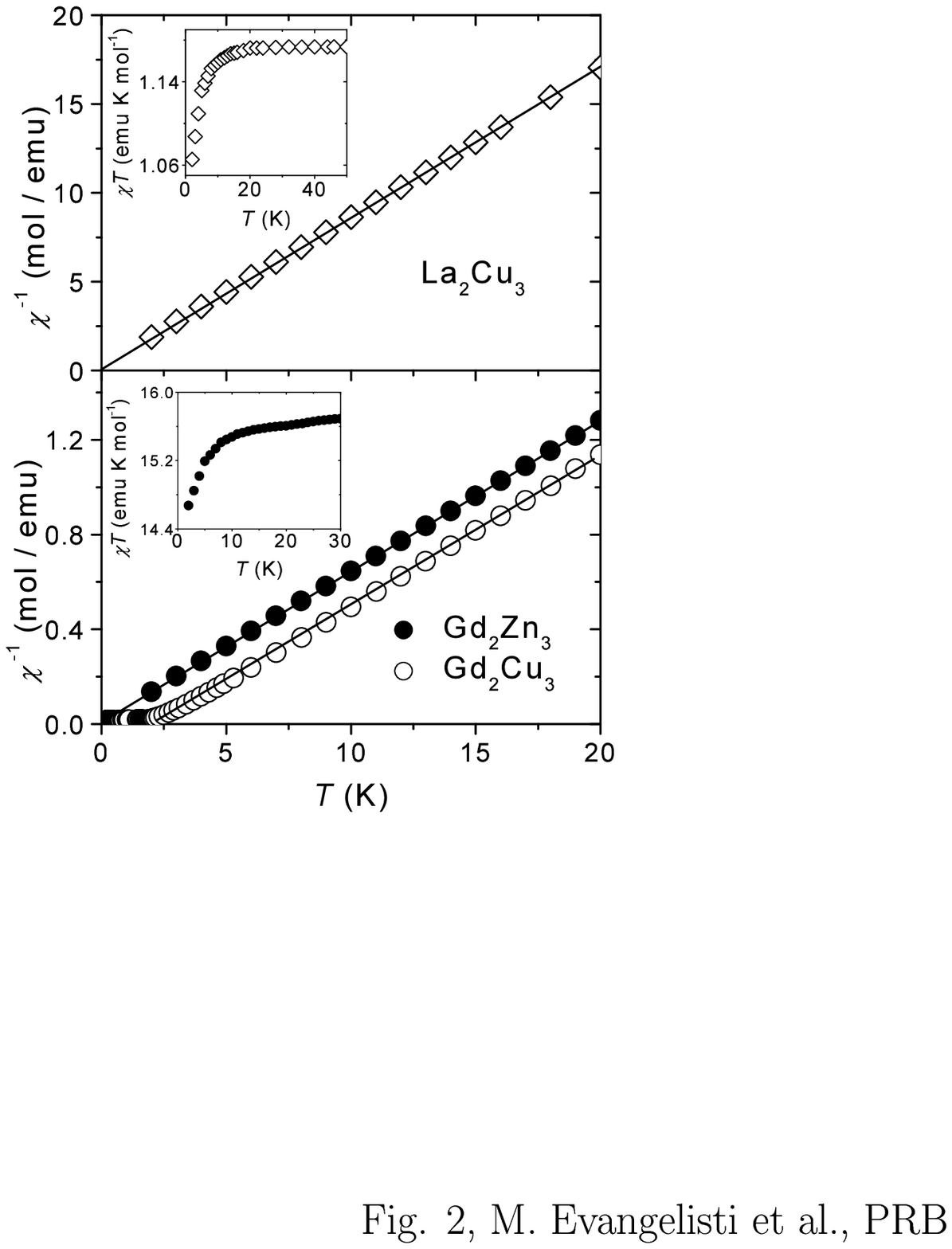}
\clearpage
\includegraphics[width=17cm]{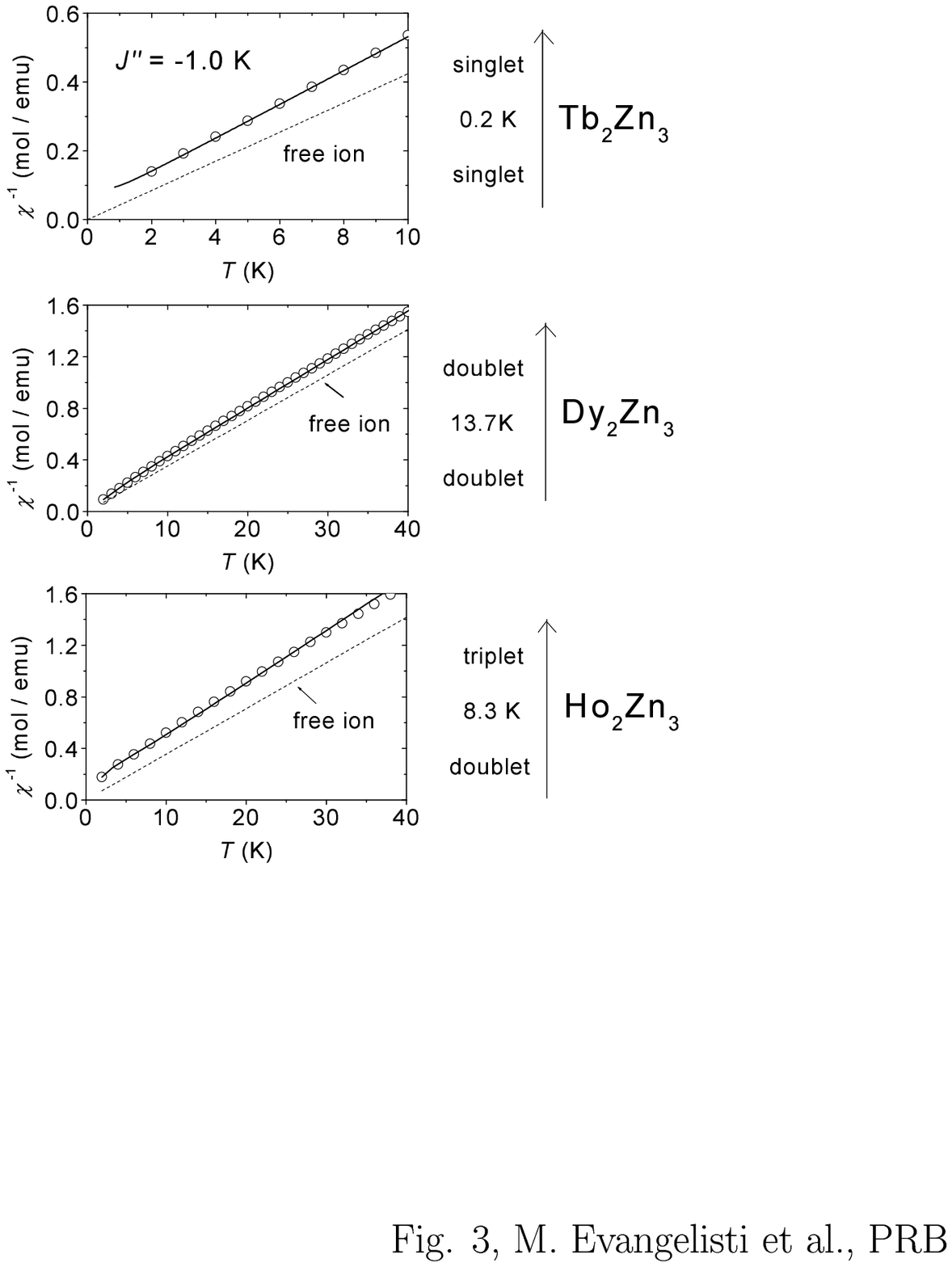}
\clearpage
\includegraphics[width=17cm]{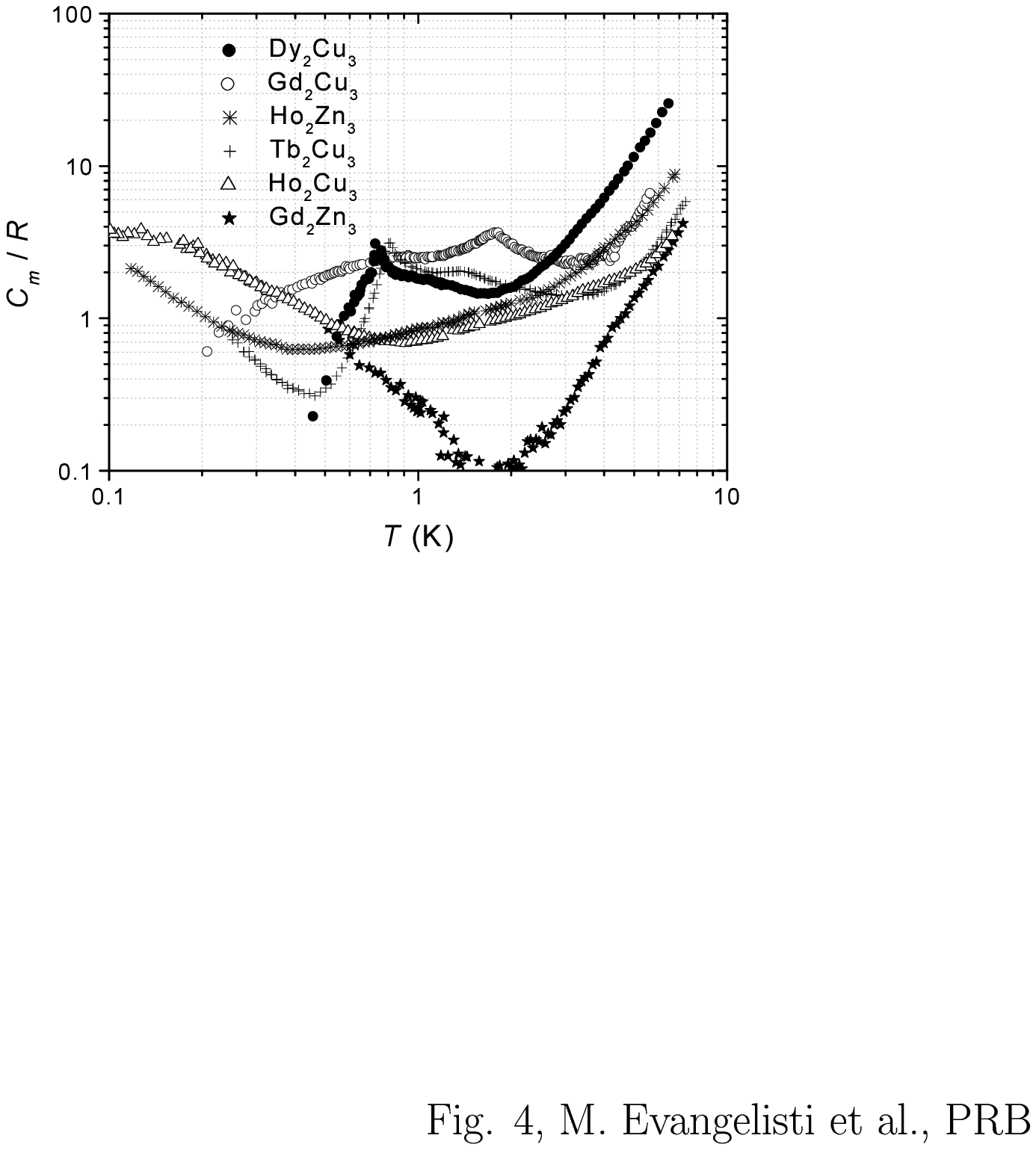}
\clearpage
\includegraphics[width=17cm]{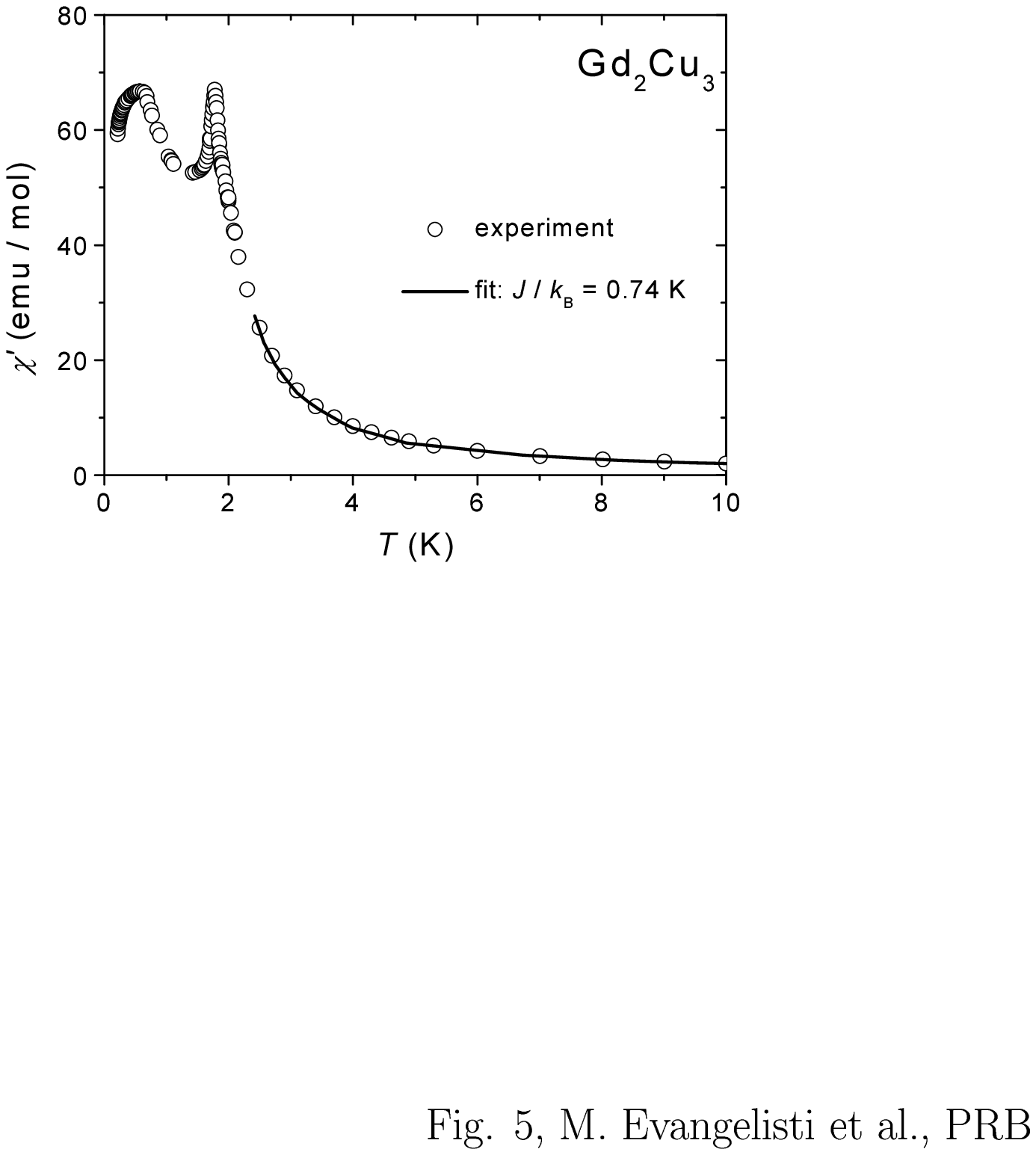}
\clearpage
\includegraphics[width=17cm]{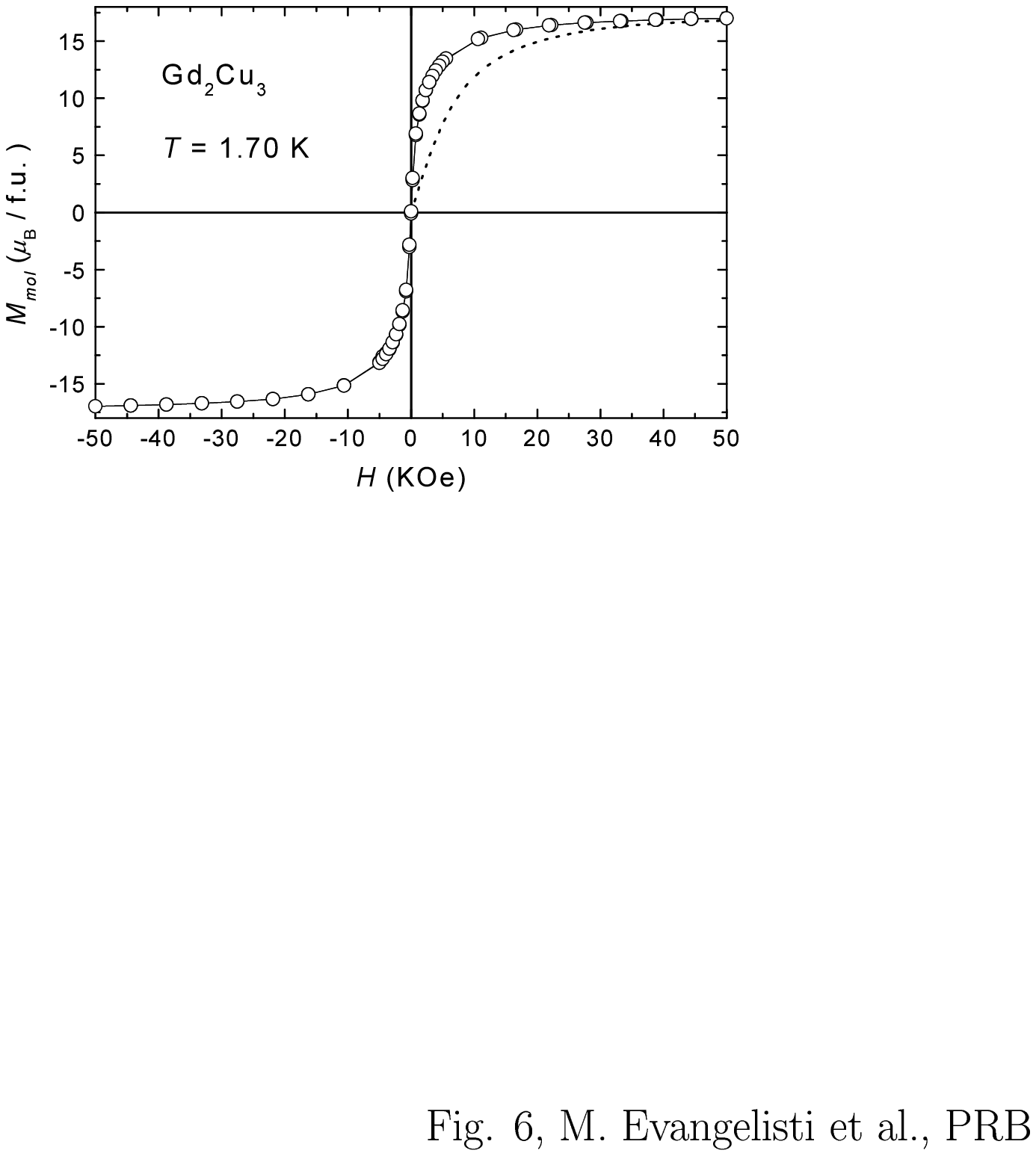}
\clearpage
\includegraphics[width=17cm]{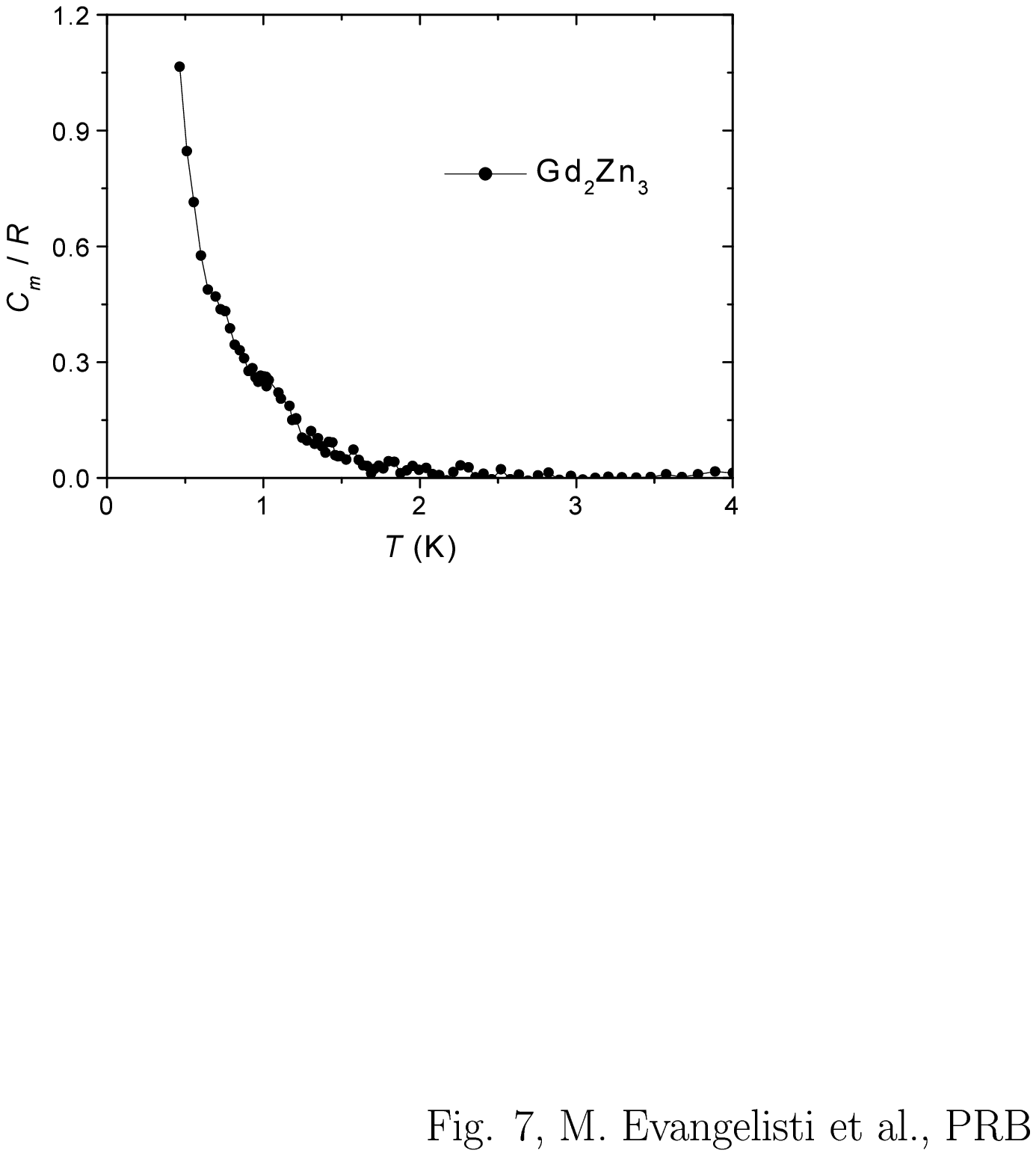}
\clearpage
\includegraphics[width=17cm]{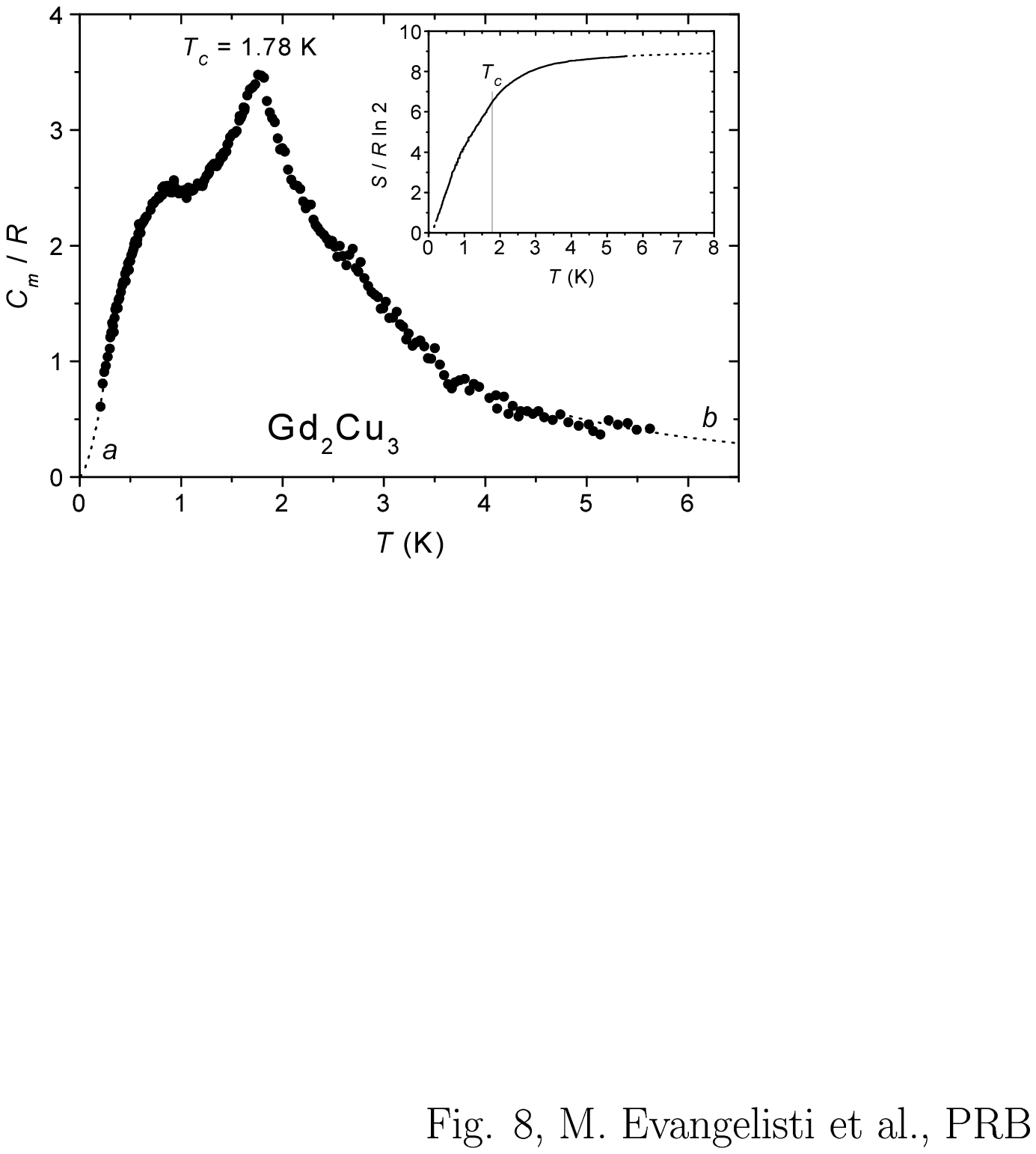}
\clearpage
\includegraphics[width=17cm]{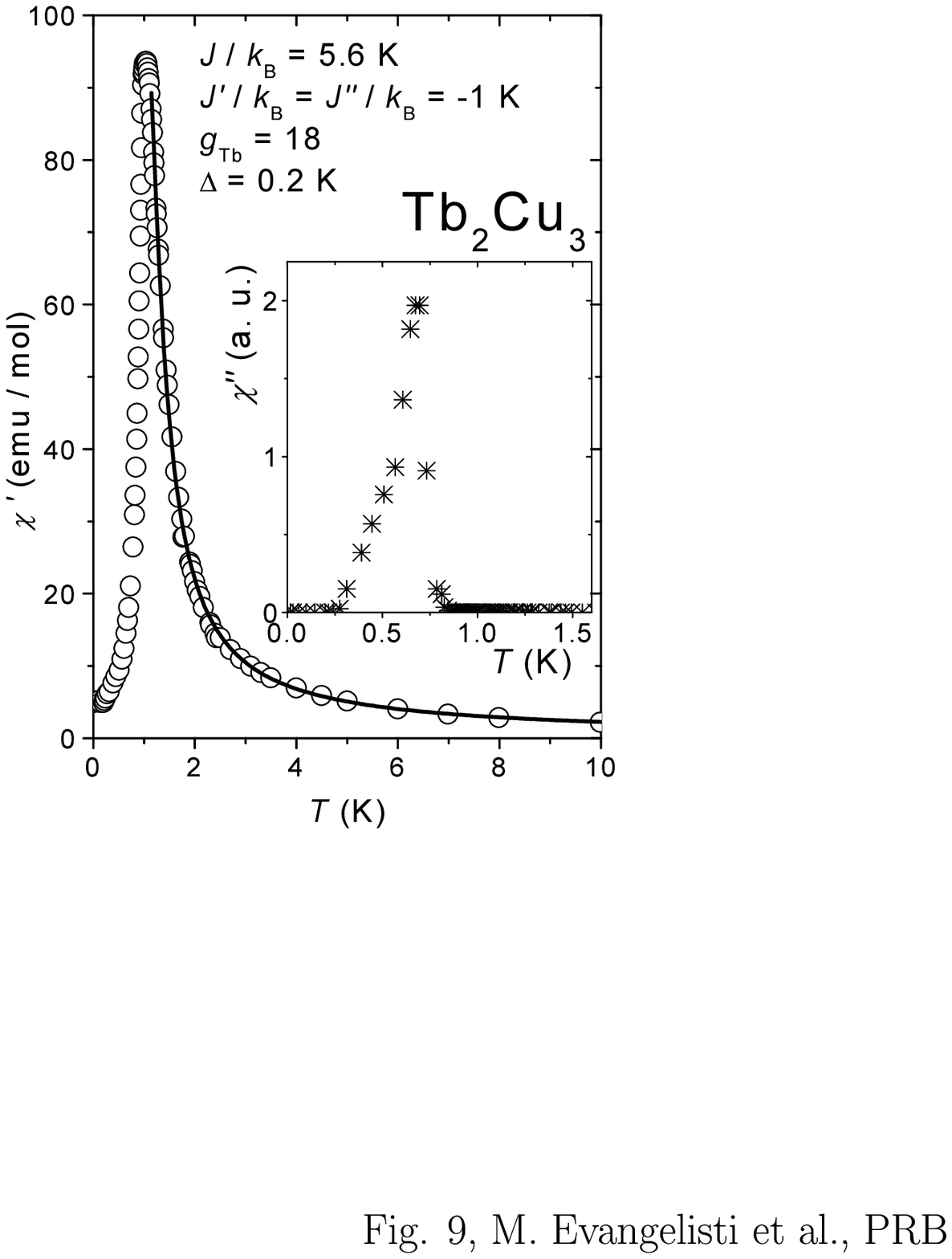}
\clearpage
\includegraphics[width=17cm]{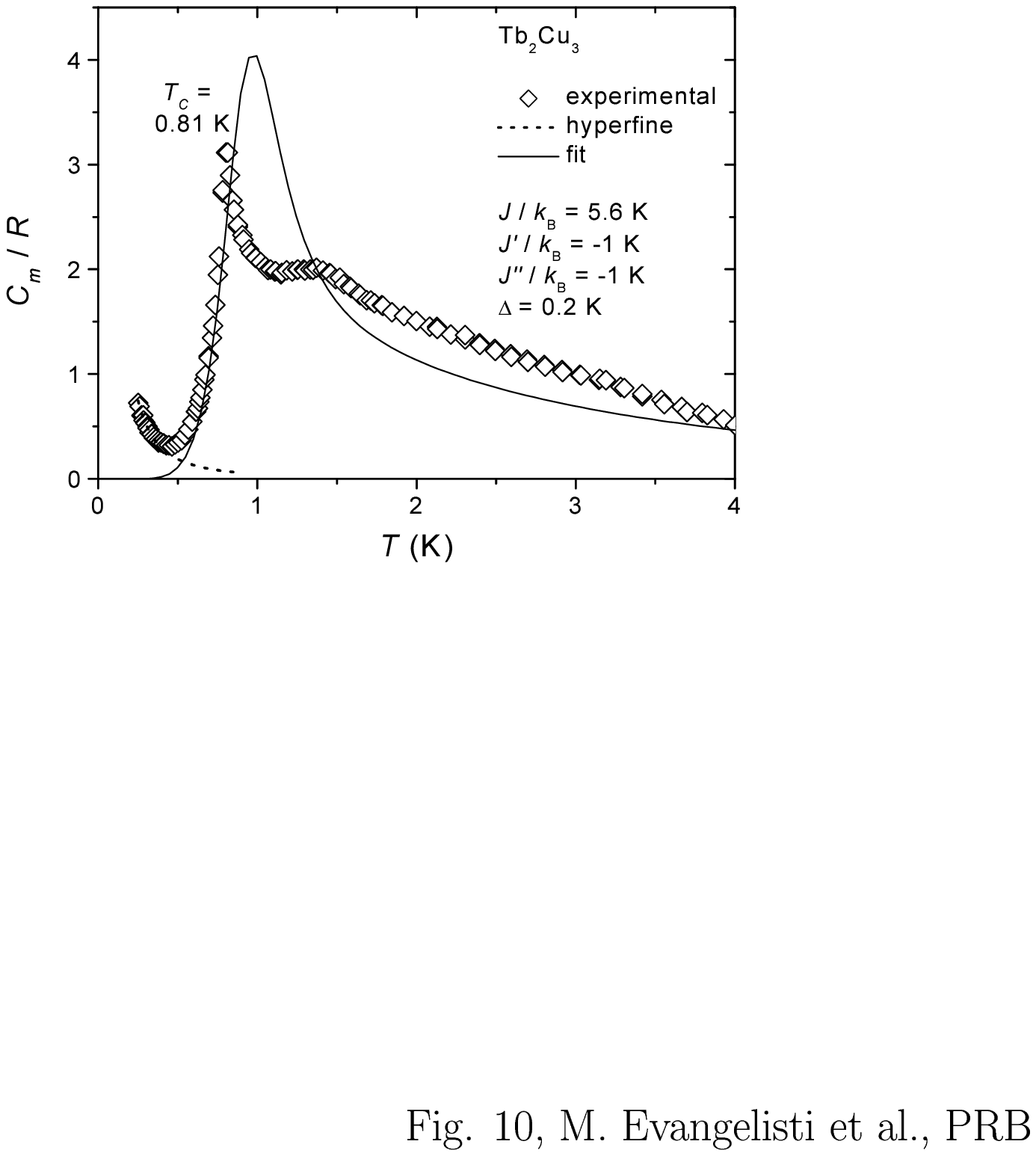}
\clearpage
\includegraphics[width=17cm]{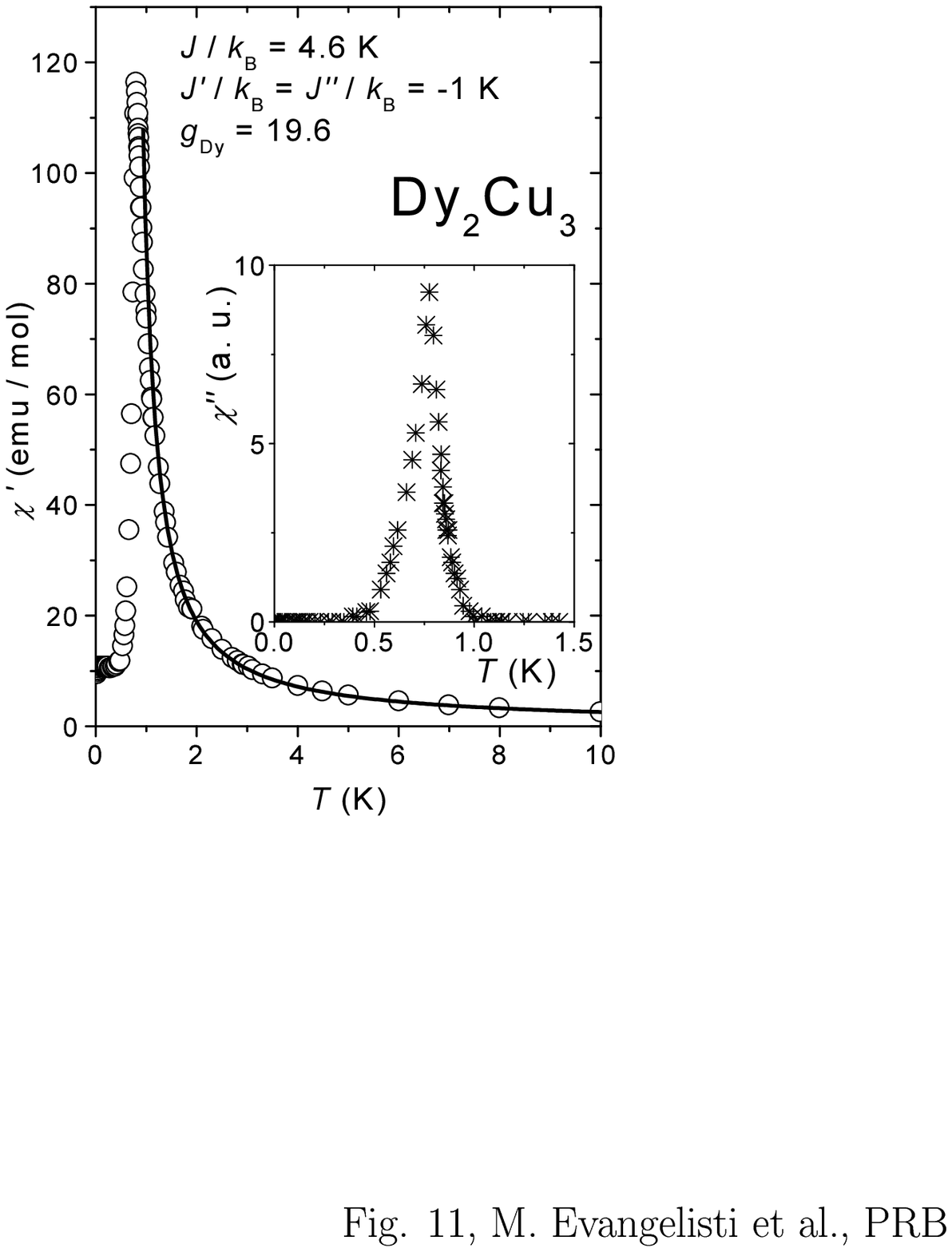}
\clearpage
\includegraphics[width=17cm]{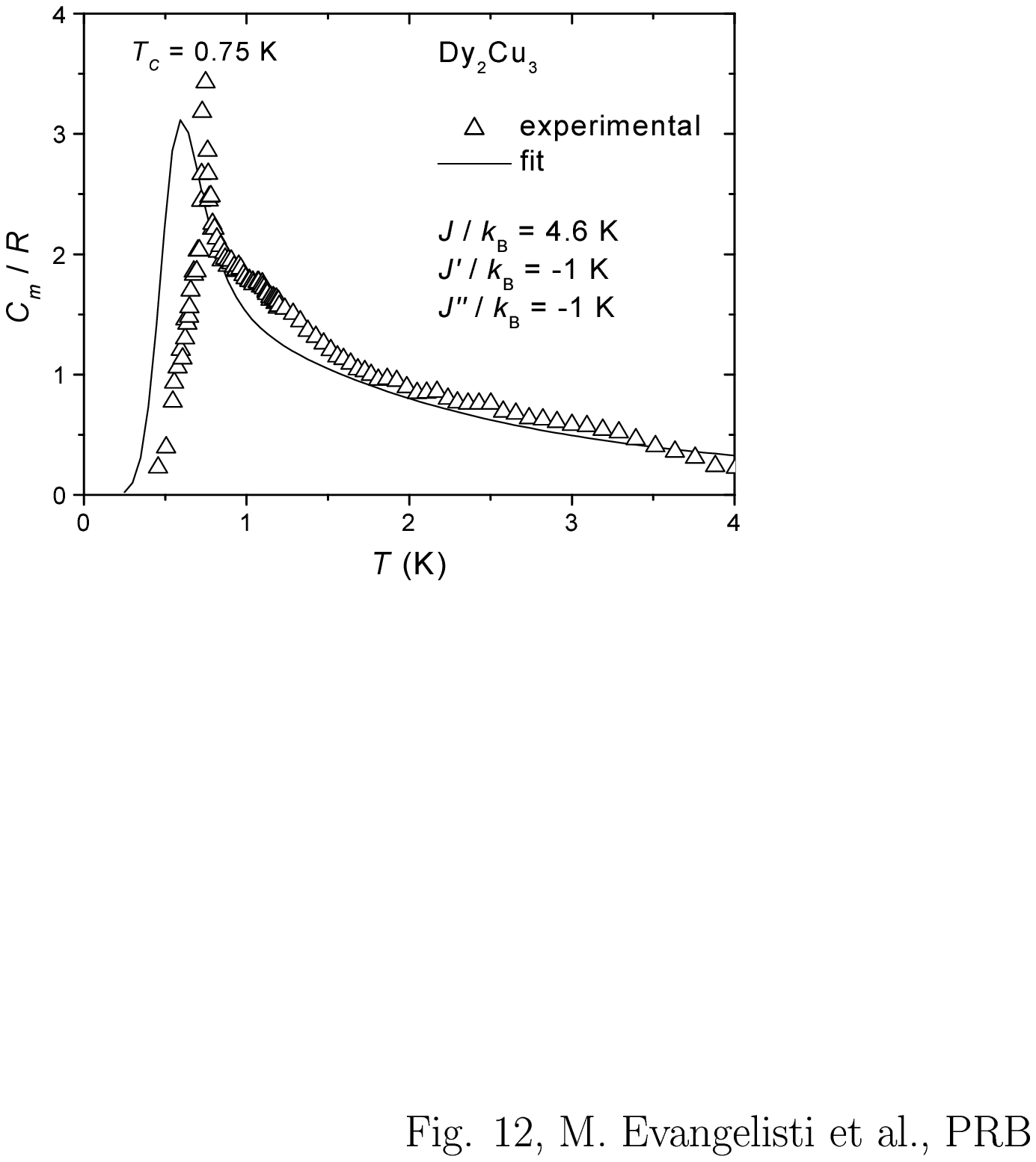}
\clearpage
\includegraphics[width=17cm]{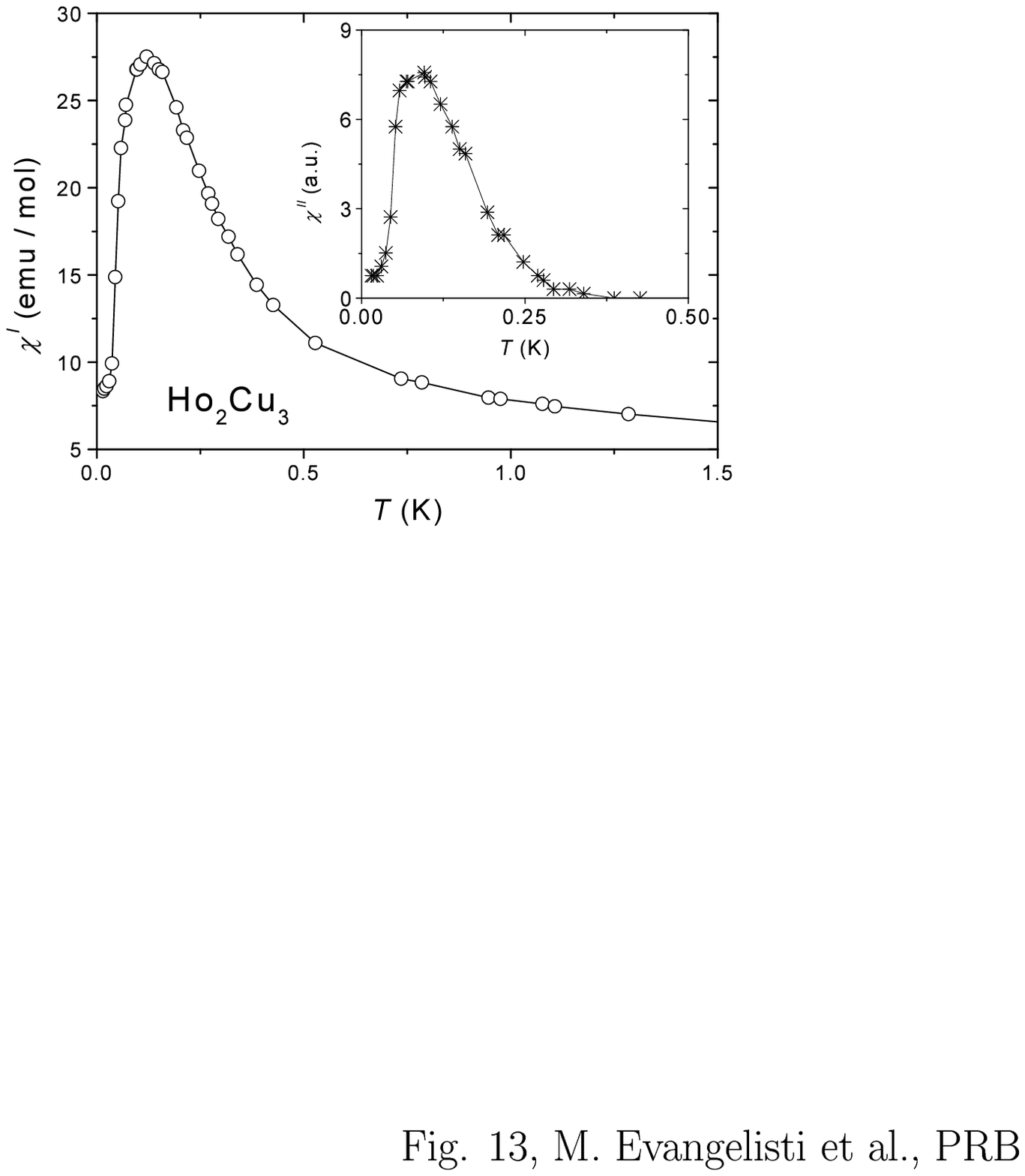}
\clearpage
\includegraphics[width=17cm]{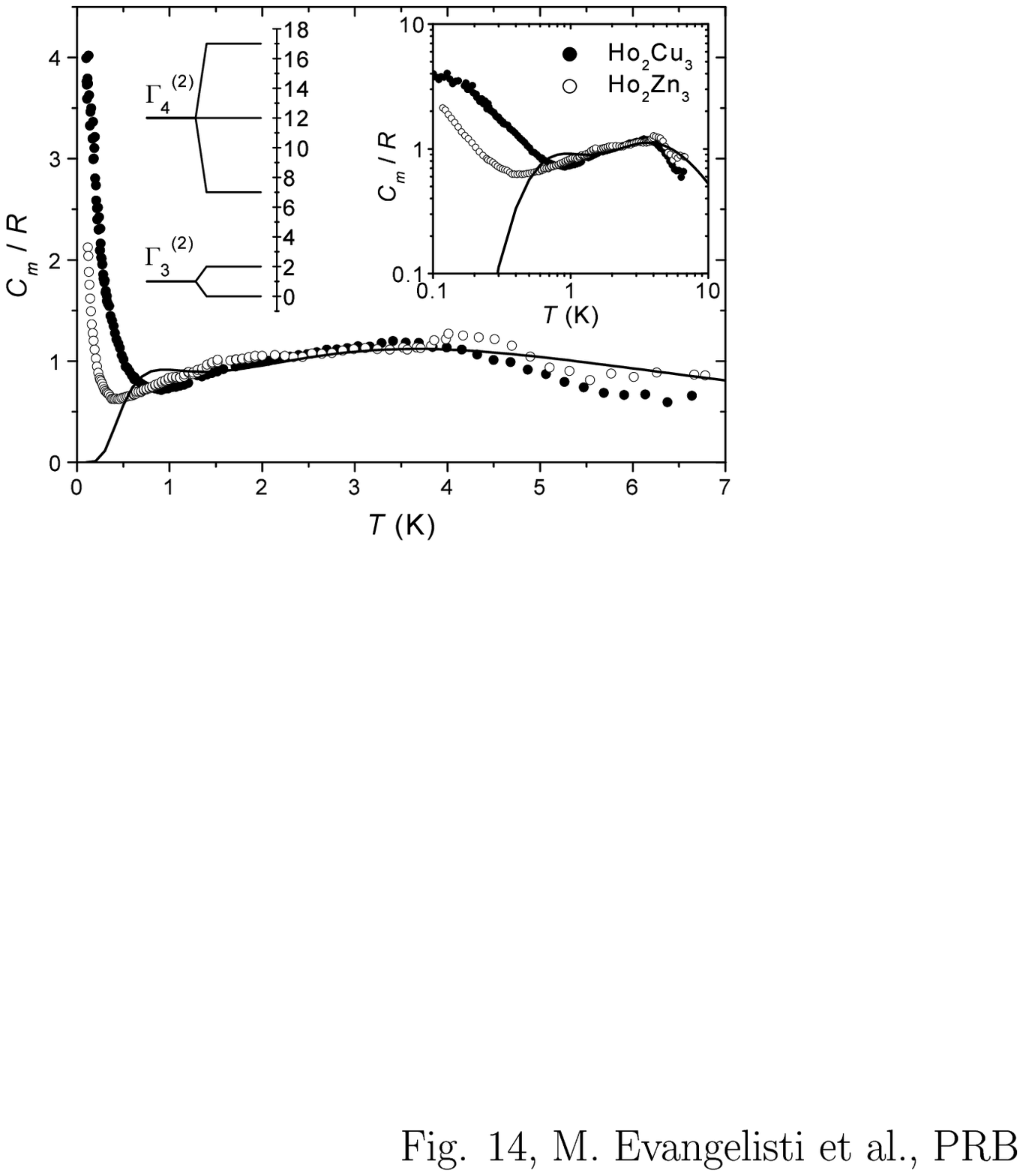}
\clearpage
\includegraphics[width=17cm]{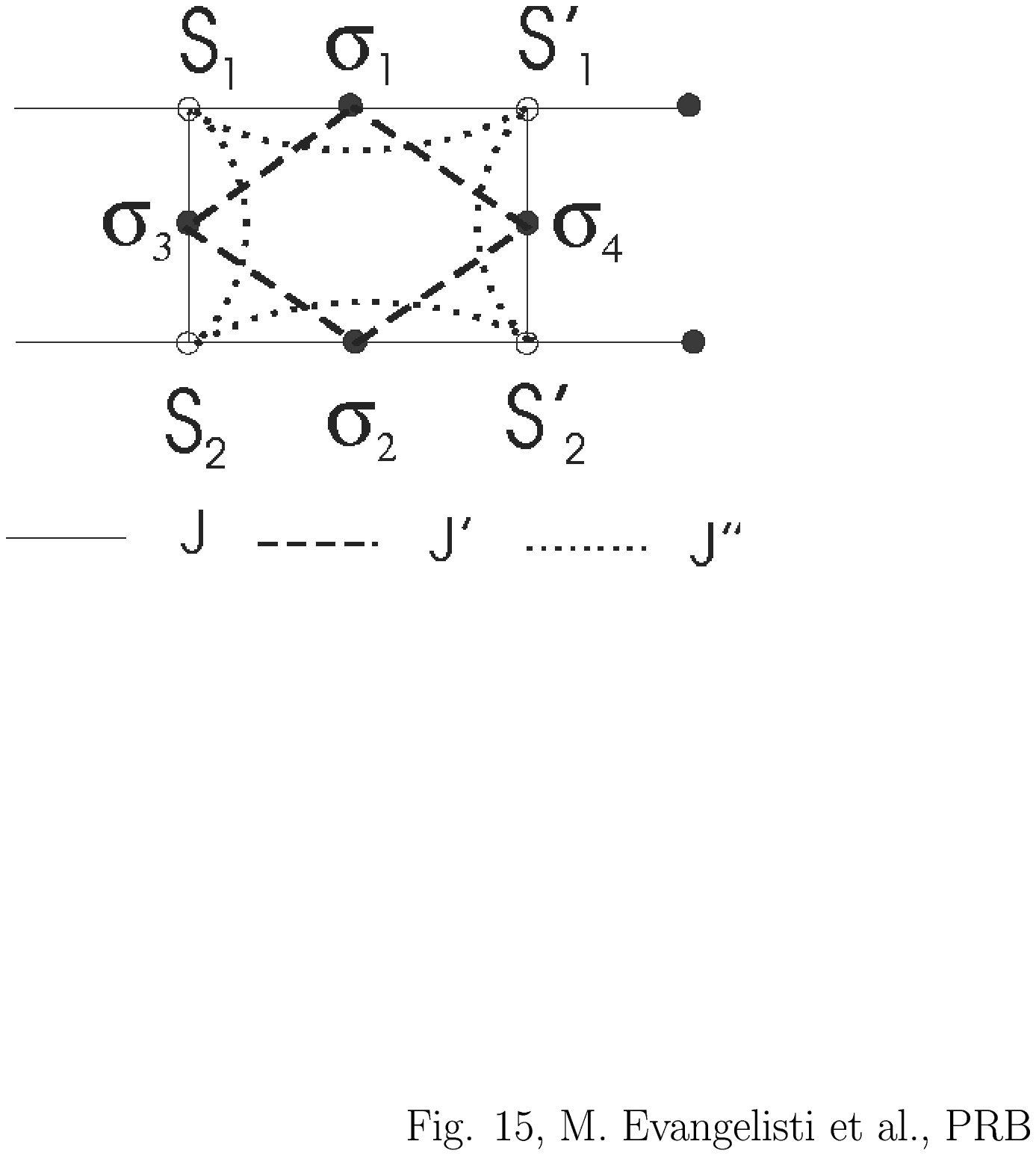}


\begin{thebibliography}{}
\bibitem{mfcurie} Considering the simple case of isotropic interaction with no
anisotropy term but with a magnetic field added, $\mathcal{H}=-J\sum_{i,j}~\vec S^{ i}\cdot\vec S^{
j}-g\mu_{{\rm B}}\sum_{i}\vec S^{ i}\cdot\vec H$, the mean field expression for the Curie-Weiss
temperature is thus given by $k_{{\rm B}}\theta=\frac{1}{3}zJS(S+1)$, where $z$ is the number of
neighboring spins.
\end{thebibliography}
\end{document}